\documentclass{emulateapj}
\usepackage{lscape}
\usepackage{amsmath}
\usepackage{amsthm}
\usepackage{amssymb}

\slugcomment{\footnotesize Accepted for Publication by ApJ.}

\newcommand{\degree}{\mbox{$^{\circ}$}}

\newcommand{\am}{\mbox{\arcmin}}
\newcommand{\as}{\mbox{\arcsec}}

\newcommand{\kms}{\mbox{km s$^{-1}$}}
\newcommand\cmv{\mbox{cm$^{-3}$}}
\newcommand\cmc{\mbox{cm$^{-2}$}}

\newcommand{\um}{$\mu$m}

\newcommand\sed{spectral energy distribution}
\newcommand{\msun}{\mbox{M$_\odot$}}

\newcommand{\tex}{\mbox{$T_{\rm ex}$}}
\newcommand{\tmb}{\mbox{$T_{\rm mb}$}}

\newcommand{\tk}{\mbox{$T_K$}}
\newcommand{\td}{\mbox{$T_D$}}

\newcommand{\np}{\mbox{$n_p$}}

\newcommand{\mvir}{\mbox{$M_{vir}$}} 
\newcommand{\miso}{\mbox{$M_{iso}$}} 
\newcommand{\mean}[1]{\mbox{$\langle#1\rangle$}} 

\newcommand{\ammonia}{\mbox{{\rm NH}$_3$}}

\newcommand{\co}{$^{12}$CO}
\newcommand{\coo}{$^{13}$CO}

\input{epsf}

\newcommand{\Snu}{\mbox{$S_{\nu}$}}

\begin{document}

\title {\bf The Bolocam Galactic Plane Survey -- III.  Characterizing Physical Properties of Massive Star-Forming Regions in the Gemini OB1 Molecular Cloud}
\author{Miranda K.~Dunham\altaffilmark{1,2}, 
Erik Rosolowsky\altaffilmark{3},
Neal J.~Evans II\altaffilmark{1},
Claudia J.~Cyganowski\altaffilmark{4},
James Aguirre\altaffilmark{5},
John Bally\altaffilmark{6},
Cara Battersby\altaffilmark{6},
Eric Todd Bradley\altaffilmark{7},
Darren Dowell\altaffilmark{8},
Meredith Drosback\altaffilmark{9},
Adam Ginsburg\altaffilmark{6},
Jason Glenn\altaffilmark{6},
Paul Harvey\altaffilmark{1,6},
Manuel Merello\altaffilmark{1},
Wayne Schlingman\altaffilmark{10},
Yancy L.~Shirley\altaffilmark{10},
Guy S.~Stringfellow\altaffilmark{6},
Josh Walawender\altaffilmark{11},
Jonathan P.~Williams\altaffilmark{12}
}

\altaffiltext{1}{Department of Astronomy, The University of Texas at Austin,
     1 University Station C1400, Austin, Texas 78712--0259}
\altaffiltext{2}{E-mail:  nordhaus@astro.as.utexas.edu}
\altaffiltext{3}{University of British Columbia, Okanagan,
     3333 University Way, Kelowna BC V1V 1V7 Canada}
\altaffiltext{4}{Department of Astronomy, University of Wisconsin, Madison, WI 53706}
\altaffiltext{5}{Department of Physics and Astronomy, University of Pennsylvania, Philadelphia, PA}
\altaffiltext{6}{CASA, University of Colorado, 389-UCB, Boulder, CO 80309}
\altaffiltext{7}{Department of Physics, University of Central Florida}
\altaffiltext{8}{Jet Propulsion Laboratory, California Institute of Technology, 4800 Oak Grove Dr., Pasadena, CA 91104}
\altaffiltext{9}{Department of Astronomy, University of Virginia, P.O. Box 400325, Charlottesville, VA 22904}
\altaffiltext{10}{Steward Observatory, University of Arizona, 933 North Cherry Ave., Tucson, AZ 85721}
\altaffiltext{11}{Institute for Astronomy, University of Hawaii, 640 N. Aohoku Pl., Hilo, HI 96720}
\altaffiltext{12}{Institute for Astronomy, University of Hawaii, 2680 Woodlawn Dr., Honolulu, HI 96822}

\begin{abstract}
We present the 1.1 millimeter Bolocam Galactic Plane Survey (BGPS)
observations of the Gemini OB1 molecular cloud complex, and targeted
\ammonia\ observations of the BGPS sources.  When paired with
molecular spectroscopy of a dense gas tracer, millimeter observations
yield physical properties such as masses, radii, mean densities,
kinetic temperatures and line widths.  We detect 34 distinct BGPS
sources above 5$\sigma=0.37$ Jy~beam$^{-1}$ with corresponding
5$\sigma$ detections in the \ammonia(1,1) transition.  Eight of the
objects show water maser emission (20\%).  We find a mean millimeter
source FWHM of 1.12 pc, and a mean gas kinetic temperature of 20 K
 for the sample of 34 BGPS sources with detections in the \ammonia(1,1) line.
The observed \ammonia\ line widths are dominated by non-thermal
motions, typically found to be a few times the thermal sound speed
expected for the derived kinetic temperature.  We calculate the
mass for each source from the millimeter flux assuming the sources are isothermal and 
find a mean isothermal mass within a 120\as\ aperture of $230\pm180$ \msun.
We find a total mass of 8,400 \msun\ for all BGPS sources in the Gemini OB1
molecular cloud, representing 6.5\% of the cloud mass.  By comparing the
millimeter isothermal mass to the virial mass within a radius equal to the
mm source size calculated from the \ammonia\ line
widths, we find a mean virial parameter (\mvir/\miso) of 1.0$\pm$0.9 for the 
sample.  We find mean values for the distributions of column densities of $1.0\times10^{22}$ \cmc\ for H$_2$, and $3.0\times10^{14}$ \cmc\ for \ammonia, giving a mean
\ammonia\ abundance of $3.0\times10^{-8}$ relative to H$_2$.  We find
volume-averaged densities on the order of $10^3-10^4$ \cmv. 
The sizes and densities suggest
that in the Gem OB 1 region the BGPS is detecting the clumps from
which stellar clusters form, rather than smaller, higher density cores
where single stars or small multiple systems form.

\end{abstract}

\section{Introduction}\label{intro}
The availability of millimeter bolometer arrays has made possible
large-scale, blind surveys of the cold dust most closely associated
with star formation (e.g., Enoch et al. 2007; Motte et al. 2007;
Aguirre et al. 2010; Schuller et al. 2009).  Such blind millimeter continuum surveys
can identify the clumps where massive stars and clusters are born
without the requirement of a signpost of massive star formation, such
as maser emission (e.g., Palla et al. 1991; Plume et al. 1992; Plume et
al. 1997; Beuther et al 2002; Mueller et al. 2002; Shirley et
al. 2003), radio continuum emission (e.g., Ramesh \& Sridharan 1997;
Sridharan 2002), infrared sources (e.g., Molinari et al 2000; Kumar
et al. 2006; Kumar \& Grave 2007; Robitaille et al 2008), or infrared
colors (Wood \& Churchwell 1989).   
Infrared dark clouds (IRDCs; regions of high density seen in absorption against the diffuse
mid-IR background which are thought to represent the earliest stages of massive star formation) provide another way to find regions of massive
star formation
(Rathborne et al. 2006; Jackson et al. 2008), but only if they lie in
front of extended mid-IR emission.  However, IRDCs are seen in emission at mm 
wavelengths, allowing the BGPS to detect the earliest stages of massive
star formation at larger distances
where a significant mid-IR background is absent or too much foreground
mid-IR emission is present.


Galaxy-wide millimeter continuum surveys, in conjunction with distance
information, will allow us to determine the physical properties of a
vast number of sources and begin answering some important questions.
What is the impact of environment on star formation?  How do the
physical properties of clumps, such as size, mass, and mean density
vary as a function of Galactocentric radius and proximity to a spiral
arm?  Answering these questions will provide a greater understanding
of star formation in our own Galaxy.  Also, by characterizing the
distribution of dense gas clumps in our own galaxy, we can better
understand the aggregate emission of clumps in distant galaxies where
the dense gas structures cannot be resolved.

In this paper, we refer to clumps as distinct bound regions within
molecular clouds that will form entire stellar clusters, and cores
as the regions within clumps that will form a single star or small
multiple system (e.g., Williams et al.~2000).  
Clumps can be characterized with masses ranging from $50-500$ \msun, 
sizes ranging from $0.3-3$ pc, mean densities ranging from $10^3 - 10^4$ 
\cmv, and gas temperatures ranging from $10-20$ K,
while starless cores can be characterized by masses from $0.5-5$
\msun, sizes of $0.03-0.2$ pc, mean densities of $10^4 - 10^5$ \cmv, and gas 
temperatures between 8 and 12 K (Bergin \& Tafalla 2007).

The Bolocam Galactic Plane Survey (BGPS) is one of the first
millimeter surveys of the northern Galactic plane (Aguirre et
al. 2009), and has cataloged 8,358 sources (Rosolowsky et al. 2009).
The BGPS consists of two parts: a blind survey of the inner Galaxy,
and a targeted survey of well-known star-forming regions in the outer
Galaxy.  The inner Galaxy survey coverage is continuous over
$-10\degree < l < 90.5\degree$, and the outer Galaxy coverage includes
IC1396, the W3/4/5 region, 4 square degrees towards the Perseus Arm
(near NGC 7538), and the Gemini OB1 molecular cloud.  With a large
range in Galactic longitude, the BGPS provides the opportunity to
study star formation in drastically different environments.  


Millimeter continuum emission detected in the BGPS can arise from
all objects along a line of sight, from a few hundred pc to the other side
of the Galaxy; distances to BGPS sources are not known a priori.
With recent improvements to kinematic models of the Galaxy
(e.g., Pohl et al. 2008) and VLBI parallax distances (Reid et
al. 2009), radial velocities can yield kinematic distances.  Molecular
spectral lines also provide information regarding the kinematics
within each source.  Ammonia (\ammonia), in particular, is an
excellent tracer of dense gas which also provides information
regarding the line of sight velocity dispersion, kinetic gas
temperature, column density and mass (e.g., Ho \& Townes 1983;
Zinchenko et al. 1997; Swift et al. 2005; Rosolowsky et al. 2008;
Foster et al. 2009).  Pairing the 1.1 mm continuum data with
\ammonia\ observations allows complete characterization of both the gas and
dust properties (e.g., Rosolowsky et al. 2008; Foster et al. 2009).

In this paper, we present the results of an \ammonia\ survey of the
BGPS sources in the Gemini OB1 molecular cloud. In
\S\ref{gemob1moleccloud} we provide an overview of the structure and
previous observations of the Gemini OB1 molecular cloud.  We briefly
describe the BGPS and \ammonia\ surveys and data reduction in
\S\ref{surveys}, and present the parameter estimation methods in
\S\ref{parameterestimation}.  In \S\ref{results} we discuss the
basic results of the BGPS and \ammonia\ in the Gemini OB1 molecular
cloud, and in \S\ref{analysis} we discuss
the derived physical properties in detail.  We compare
our results to previous studies in \S\ref{comparison}, and explore the 
effects of distance on the derived properties in \S\ref{bgpssource}.  
Finally, we present a summary in \S\ref{summary}.  In future work, we 
will present GBT+BGPS observations of sources at low Galactic longitudes.

Studying the Gem OB1 region is particularly useful because the BGPS
sources in this region are located in the same molecular complex and
thus have similar distances (Carpenter et al. 1995a; Carpenter
et al. 1995b).  We are able to investigate the physical properties of
this subset of BGPS sources without the ambiguity introduced by a
large range of distances seen for the sample as a whole.  
This paper will serve as a Galactic anti-center comparison for a study of
\ammonia\ observations of BGPS sources at a range of Galactic 
longitudes in the inner Galaxy (M.~K.~Dunham et al., in prep.).


\section{Gemini OB1 Molecular Cloud}\label{gemob1moleccloud}

Gemini OB1 is an OB association centered near $(l,b)=(189$\degree$,1$\degree$)$ which was first identified by Morgan, Whitford, \& Code (1953).  It became known as Gem OB1 when
Humphreys (1978) compiled a list of the brightest OB stars in the Milky Way, cataloging
20 OB stars as belonging to the association.  Many groups have studied the 
association and found distances ranging from 1.4 kpc to 1.9 kpc (Crawford et al. 1955; 
Hardie et al. 1960; Georgelin et al. 1973; Humphreys 1978).  There is also a significant spread
in age amongst the association members, with ages ranging from 2 Myr (Grasdalen \&
Carrasco 1975) to 10 Myr (Barbaro et al. 1969).

The Gemini OB1 molecular cloud was first identified as an enhancement
of emission in the Columbia \co\ survey of the Galactic plane (Dame et
al. 1987), and spans a $6\degree\times6\degree$ area on the sky (Huang
\& Thaddeus 1986).  There are many star-forming regions in the Gem OB1
molecular cloud, including well-known HII regions S247, S252, the
group S254, S255, S256, S257, and S258 (hereafter referred
to as S254-258), S259, S261 (Sharpless 1959), and BFS 52 as well
as a supernova remnant, IC 443.  Reid et al. (2009) used the VLBA to
measure the parallax of methanol masers in S252, and found a
distance of $2.10\pm0.026$ kpc, which we adopt in this study.

Studies of \co\ and \coo\ in the Gem OB1 molecular cloud show abundant
arc and ring-like filamentary structure (Carpenter et al. 1995a;
K\"ompe et al. 1989).  Carpenter et al. (1995a) mapped 32 square
degrees of the Gem OB1 molecular cloud in \co($J=1-0$) and
\coo($J=1-0$) with 50\as\ sampling and 45\as\ and 50\as\ resolution
respectively.  They suggest that the filamentary structure is
comprised of molecular material which has been swept-up by the
expanding HII regions and wind blown bubbles.  They conclude that the
high column density regions where star formation is currently
occurring were formed through the interaction of newly formed stars
and the surrounding molecular material.  Carpenter et al. (1995a)
split the cloud into 7 separate regions and find that 3 regions (S247, S252, and S254-258)
contain active star formation based on higher H$_2$ column densities
($> 10^{22}$ \cmc) and warmer kinetic temperatures (10 to 30 K).  

Carpenter et al.~(1995b) identified 11 clumps within the
\co\ filaments using CS($J=2-1$) observations (55\as\ resolution), all
with masses greater than 100 \msun.  They found that at least eight of
the 11 clumps have associated IRAS point sources.  Carpenter et
al. (1995b) also conducted a near-infrared imaging survey of a
$30\am\times45\am$ area surrounding S247, including three of the
CS clumps.  They found that each of the three clumps contains a
cluster of stars in the near-infrared images.

Chavarr\'ia et al.~(2008) identified 510 young stellar objects (YSOs)
with near or mid-IR excess in the S254-258 complex using a
combination of \emph{Spitzer} IRAC and near-infrared data.  They
classify 87 Class I sources, and 165 Class II sources, and find that
80\% of the YSOs are in clusters surrounded by a more evenly
distributed population.

\begin{figure}
\begin{center}
\includegraphics[angle=0,width=3.8 in]{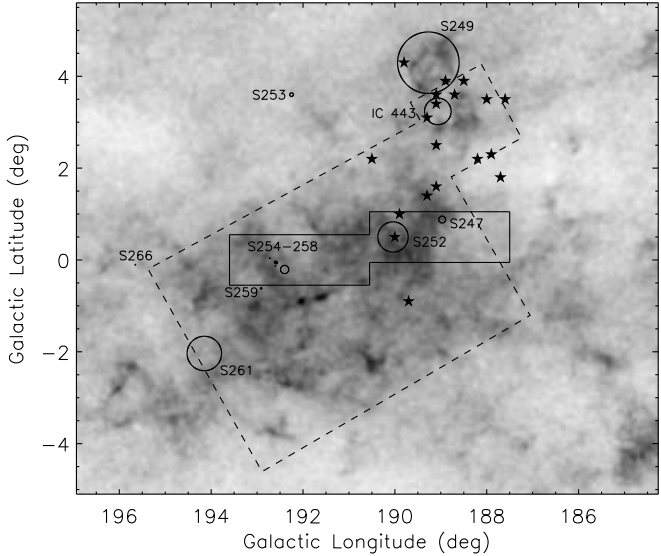}\caption{\label{gemob1cloud}
 The Gemini OB1 molecular cloud.  Extinction maps of Dobashi et
 al. (2005) are shown in inverted grayscale ranging from -0.5 to 4.0 magnitudes.  
 The OB association
 members (Humphreys 1978) are shown as stars.  The HII regions
 (Sharpless 1959) and IC 443 are shown as circles sized to match the
 extent of the region.  The dashed line marks the area mapped in
 \co\ and \coo\ by Carpenter et al. (1995a), and the solid line marks
 the extent of the 1.1 mm BGPS maps.  }
\end{center}
\end{figure}

Figure \ref{gemob1cloud} shows the 
extinction maps of Dobashi et al. (2005) for the Gemini OB1 molecular cloud.
OB association members are shown as stars (Humphreys 1978), HII regions
(Sharpless 1959) and IC 443 are shown as circles with sizes corresponding to 
the source size.  The extent of the \co\ and \coo\ maps of Carpenter et al. (1995a)
is marked by the dashed line, while the solid line shows the extent of the BGPS
map presented in this paper.

\section{Surveys}\label{surveys}
\subsection{Bolocam 1.1 mm Galactic Plane Survey}\label{bgps}

The BGPS\footnote{See http://milkyway.colorado.edu/bgps/.} has
surveyed approximately 170 square degrees of the Galactic Plane using
the Bolocam instrument on the Caltech Submillimeter
Observatory\footnote{The Caltech Submillimeter Observatory is
 supported by the NSF.} (CSO) on Mauna Kea. Aguirre et al.~(2010)
present specific details regarding the BGPS and methods and
Rosolowsky et al.~(2010) detail the BGPS catalog.  The images and
catalog have been made public and are hosted by the Infrared Processing
and Analysis Center via the NASA/IPAC Infrared Science
Archive\footnote{See
 http://irsa.ipac.caltech.edu/data/BOLOCAM\_GPS/.}  (IPAC).
\subsubsection{Observations}\label{bgpsobs}

\begin{figure*}
\begin{center}
\includegraphics[angle=0]{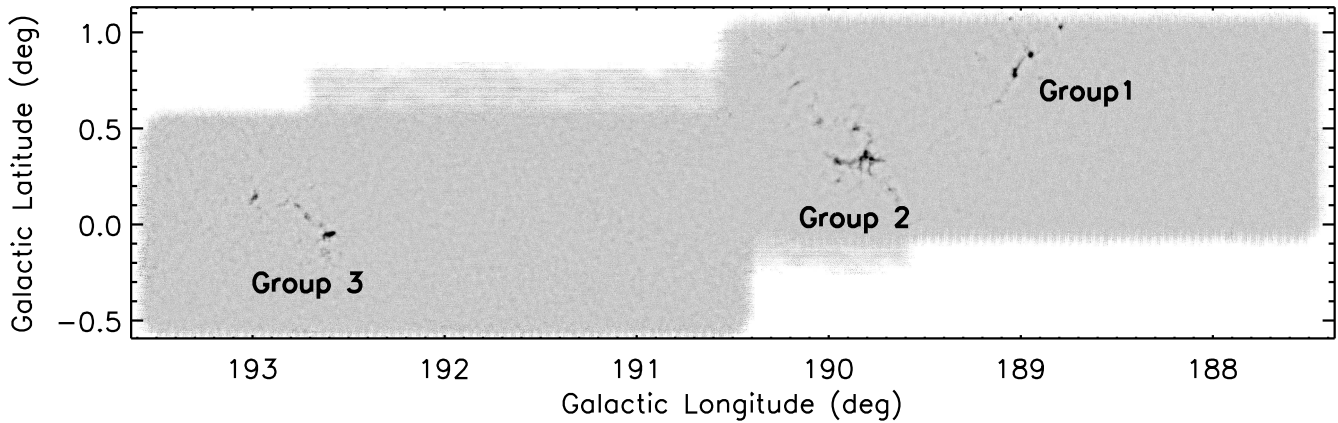} 
\end{center}
\caption{\label{largeimage} The
 BGPS 1.1 mm continuum image of the Gemini OB1 molecular cloud shown in an
 inverted, linear grayscale ranging from -0.2 to 0.75 Jy beam$^{-1}$.  Groups
 1, 2, and 3 are shown in more detail in following figures.  }
\end{figure*}

The Gem OB1 region was observed in two 3$^{\circ} \times $1$^{\circ}$
groups of observations, centered at $(l,b)=(189\degree,0.5\degree)$
and $(192\degree,0\degree)$, with a shallow observation centered
between the two main groups to aid in mosaicing the images (see Figure
\ref{largeimage}).  The centers were selected to place the most active
star-forming regions (S247, S252, and S254-258) within our
mapped region.  Observations were carried out at the Caltech
Submillimeter Observatory (CSO) as part of the BGPS between 9 and 13
September 2007.  The survey used the Bolocam instrument\footnote{See
  http://www.cso.caltech.edu/bolocam/.}, which has a hexagonal array
of 144 bolometers.  The Bolocam filter is centered at 268 GHz with a
bandwidth of 45 GHz, and was designed to exclude the $^{12}$CO(2
$\rightarrow$ 1) emission line.  Each bolometer beam is well
approximated by a Gaussian with a FWHM of 31\as, and the bolometer
separation results in an instantaneous field of view of 7.5\am.  For
the BGPS, the effective beam size after processing many scans
is 33\as\ for the current release (Aguirre et al. 2010).  For more
instrument specific details, see Glenn et al. (2003) and Haig et
al. (2004).  Observations were obtained using the raster scan mode,
with alternating scans directed along $l$ and $b$.  The scan speed was
120\as\ s$^{-1}$, and chopping was not utilized in these observations,
thereby retaining some sensitivity to large scale structure up to the
angular size of the array (see Aguirre et al. 2010 for more detailed
analysis).

\subsubsection{Reduction}\label{bgpsreduction}
At millimeter wavelengths, the detected signal is dominated by the
atmosphere, making the sky cleaning algorithm extremely important.
Bolocam's individual bolometer beams overlap substantially in the
lower atmosphere such that they probe similar columns of atmosphere,
facilitating sky subtraction.  The iterative mapping method utilizes
principal component analysis (PCA) and considers all observations for
a given field simultaneously.  In our iterative mapping method, the
temporally correlated signal is first subtracted from the time
streams.  This subtraction removes both the atmosphere and all
astrophysical structure larger than the array field of view.  The
remaining time stream, which in principle should contain only
astrophysical signal, is then mapped with
7.2\as$\times$7.2\as\ pixels, deconvolved from the beam, and returned
to a time stream.  The time stream containing only astrophysical
signal is subtracted from the original to produce a third time stream
(summing the subtracted astrophysical signals to make a source map)
that represents an estimate of the noise.  The above procedure is then
repeated for 50 iterations on the deconvolved astrophysical signal
time stream to produce the final map.  This procedure can be
fine-tuned to balance the retention of large-scale structure against
production of artifacts (such as negative bowls) around bright
sources.  The iterative mapping method resulted in a 1-$\sigma$ RMS 
noise of 0.070 Jy/beam in the Gemini OB1 image.
For more specific information on the BGPS reduction, see
Aguirre et al. (2010).

\subsubsection{Images}\label{mmimagesandcatalog}
\begin{figure}
\includegraphics[angle=0,width=3.7in]{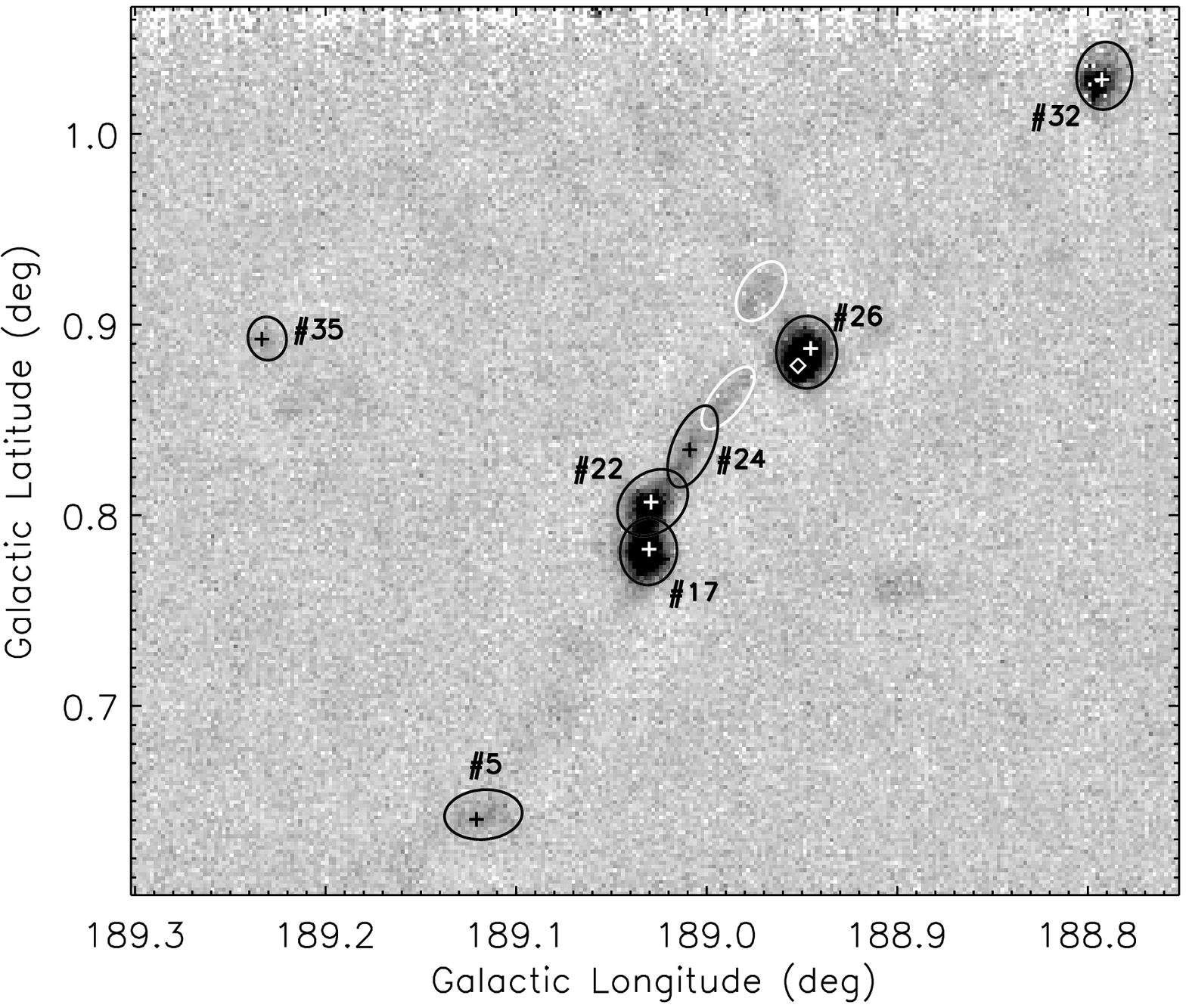} \caption{\label{group1image} The
 BGPS image of the region marked Group 1 in Figure \ref{largeimage}
 in inverted grayscale ranging from -0.2 to 0.75 Jy beam$^{-1}$.
 The BGPS sources with \ammonia\ spectra are shown as black ellipses, while
 BGPS sources without \ammonia\ spectra are shown as white ellipses.  The
 \ammonia\ pointings are shown as crosses and diamonds.  Crosses
 denote pointings closest to the peak of the millimeter emission,
 while diamonds denote additional \ammonia\ pointings within each
 BGPS source.}
\end{figure}

\begin{figure}
\includegraphics[angle=0,width=3.7in]{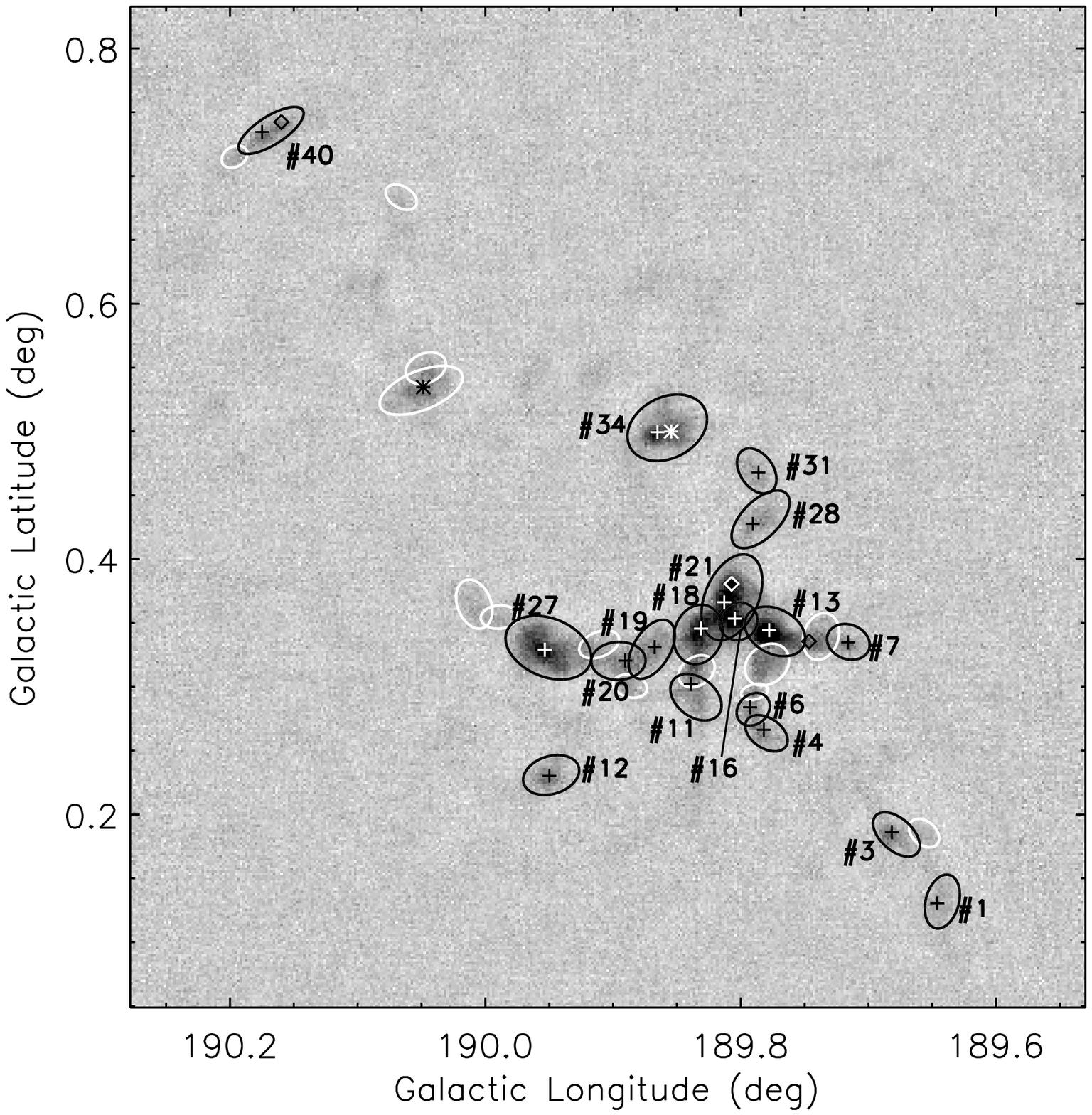} \caption{\label{group2image} The
 BGPS image of the region marked Group 2 in Figure \ref{largeimage}
 in inverted grayscale ranging from -0.2 to 0.75 Jy beam$^{-1}$.
 The BGPS sources with \ammonia\ spectra are shown as black ellipses, while
 BGPS sources without \ammonia\ spectra are shown as white ellipses.  The
 \ammonia\ pointings are shown as crosses, diamonds and asterisks.  Crosses
 denote pointings closest to the peak of the millimeter emission,
 diamonds denote additional \ammonia\ pointings within each
 BGPS source, and asterisks denote non-detections.  
 Source ellipses can overlap since these are the moment
 approximations to source sizes rather than the boundaries
 established by the extraction algorithm.}
\end{figure}

\begin{figure}
\includegraphics[angle=0,width=3.7in]{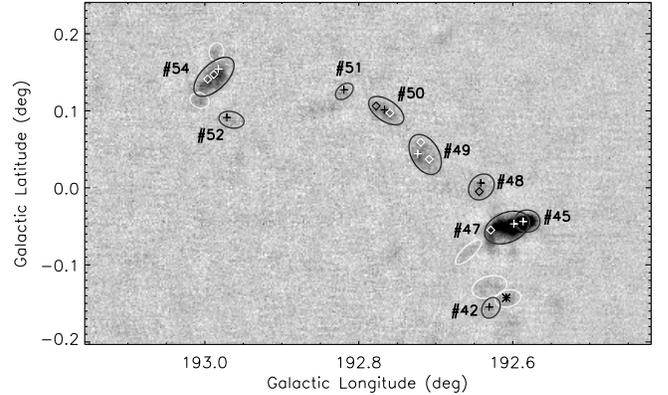} \caption{\label{group3image} The
 BGPS image of the region marked Group 3 in Figure \ref{largeimage}
 in inverted grayscale ranging from -0.2 to 0.75 Jy beam$^{-1}$.
 The BGPS sources with \ammonia\ spectra are shown as black ellipses, while
 BGPS sources without \ammonia\ spectra are shown as white ellipses.  The
 \ammonia\ pointings are shown as crosses, diamonds and asterisks.  Crosses
 denote pointings closest to the peak of the millimeter emission,
 diamonds denote additional \ammonia\ pointings within each
 BGPS source, and asterisks denote non-detections.  
 Source ellipses can overlap since these are the moment
 approximations to source sizes rather than the boundaries
 established by the extraction algorithm.}
\end{figure}

The iteratively mapped BGPS image is shown in Figure \ref{largeimage}.
The millimeter sources in our maps are highly clustered and are found
in three groups, while the remainder of the image is strikingly devoid
of millimeter emission.  This clustering is in contrast to the inner Galaxy
regions also included in the BGPS, where a similarly sized image would
contain hundreds of sources [see Aguirre et al. (2010) and J. Bally et
 al. (2010)].  The three groups correspond to S247
(referred to as Group 1; Figure \ref{group1image}), S252 (referred to
as Group 2; Figure \ref{group2image}), and S254$-$258 (referred to as
Group 3; Figure \ref{group3image}).  


\subsection{\ammonia\ Survey of BGPS Sources}\label{nh3survey}

The \ammonia\ pointings are displayed in Figures
\ref{group1image}$-$\ref{group3image} as crosses, diamonds, and asterisks.  
Since
the \ammonia\ pointings were chosen prior to completion of the BGPS catalog,
they do not necessarily correspond to the peak positions of the millimeter
emission determined by the source extraction algorithm 
(see \S\ref{mmparameterestimation}).  Crosses mark the positions closest to the peak of the millimeter emission,  
diamonds mark extra \ammonia\ pointings within each source, and the asterisks
mark the positions of \ammonia\ pointings which correspond to a mm 
source but do not have a detection in any of the \ammonia\ transitions.

\subsubsection{Observations}\label{nh3obs}
We observed 56 \ammonia\ pointings in the Gem OB1 region using the
Robert F. Byrd Green Bank Telescope\footnote{The GBT is operated by
 the National Radio Astronomy Observatory, which is a facility of the
 National Science Foundation, operated under cooperative agreement by
 Associated Universities, Inc.} (GBT) on 14 and 15 February 2008
 in two separate blocks totaling 3.6 hours.  The 23 GHz zenith
opacity for the Feb. 14 observations was $\tau_{23}\sim0.04$; and for
Feb. 15, $\tau_{23}\sim 0.11$.  The observations were broken into two
minute integrations of individual targets.  We used Feed 1 of the
high-frequency K-band receiver on the GBT as the front end for the
observations. The GBT spectrometer served as the back end, which we
configured to observe both polarizations of the feed with 50 MHz
bandwidth for four separate IFs.  The total bandwidth was sampled in
8192 channels for a 6.1 kHz ($\sim$0.08 km s$^{-1}$) channel width.
We used two different sets of IFs.  All scans contained the
NH$_3$(1,1) (23.6944955(1) GHz) and NH$_3$(2,2) (23.7226333(1) GHz)
lines.  Observations on February 14 (10 pointings) also included the
C$_2$S (N$_{J} = 2_1\to 1_0$, 22.344033(1) GHz, Yamamoto et al. 1990)
and NH$_3$(4,4) (24.1394169(1) GHz) lines.  All NH$_3$ frequencies
were taken from Lovas \& Dragoset (2003).  There were no significant
detections of the C$_2$S line and only one marginal detection of the
NH$_3$(4,4) line, so the spectrometer configuration was changed for
the following days' observations.  The observations on February 15
included the H$_2$O (22.344033 GHz), and NH$_3$(3,3) (23.8701279(5) GHz)
transitions in the remaining two IFs.  Ten pointings were observed on
14 February 2008 with C$_2$S and \ammonia(4,4), and all 56 pointings
were observed with H$_2$O and \ammonia(1,1), (2,2) and (3,3).  Ten pointings 
were observed in all six lines (four \ammonia\ lines, H$_2$O, and C$_2$S).
We observed
each source with a symmetric, in-band frequency switch (5 MHz throw).

Observations of the Gem OB1 region were conducted immediately
following observations for a pointing model correction on each night.
We calibrated the data with the injection of a noise signal
periodically throughout the observations. Because of slow variations
in the power output of the noise diodes and their coupling to the
signal path, we measured the strength of the noise signal through
observations of a source with known flux (the NRAO flux calibrator 3C
48, $S_\nu=1.13$ Jy). We repeated the flux calibration observations 
during every
observing session to detect any changes in the calibration sources,
finding no significant variations over the course of our run
($<10\%$). The beam size of the GBT is 31\as\ at these frequencies,
which projects to a size of 0.32 pc at the distance of Gem OB1 (2.10
kpc, see \S\ref{vlsr_and_d}; Reid et al. (2009)).  The GBT
\ammonia\ observations are well matched to the BGPS observations since
the GBT beam size is only a few arcseconds smaller than the effective
BGPS beam size.

\subsubsection{Data Reduction}\label{nh3reduction}
The ammonia observations were reduced using the GBTIDL reduction
package.  We extracted each scan, using the package to fold the 
spectrum to remove the effects of the
frequency switching and establish the data on the $T_A$ scale by  
calibrating the noise diodes. We
subtracted a linear baseline from the spectrum from each IF with the
exception of the IF containing the water and C$_2$S lines where a
second order baseline was required.  We then
scaled to the $T_A^*$ scale using estimates of the atmospheric opacity
interpolated in the frequency and time domains
at 22 and 23 GHz from models of the atmosphere derived using weather
data.\footnote{See http://www.gb.nrao.edu/\~ rmaddale/Weather} 
To reach the $T_{mb}$ scale, the spectra are divided by the main
beam efficiency of the GBT, which is $\eta_{mb}\sim 0.81$ at these
frequencies.  We averaged
together both polarizations; if a source was observed multiple
times, each observation was calibrated separately and the resulting
spectra were then averaged together.  The mean and standard deviation
of the RMS noise per channel for each of the 6 observed lines are listed 
in Table \ref{spectrarms}.  

\begin{deluxetable}{lcc}
\tabletypesize{\scriptsize}
\tablewidth{0pt}
\tablecaption{\label{spectrarms}Mean RMS for Spectral Line Observations}
\tablehead{
\colhead{} & \colhead{mean RMS } & \colhead{stdev RMS} \\
\colhead{Line} & \colhead{(mK)} & \colhead{(mK)}
}
\startdata
\ammonia(1,1) & 82 & 21 \\
\ammonia(2,2) & 81 & 21 \\
\ammonia(3,3) & 97 & 11 \\
\ammonia(4,4) & 52 & 1 \\
H$_2$O & 172 & 26 \\
C$_2$S & 67 & 4 
\enddata \\
\end{deluxetable}

\section{Parameter Estimation}\label{parameterestimation}
\subsection{BGPS Parameter Estimation}\label{mmparameterestimation}

The BGPS millimeter source properties are extracted from the iterative
maps using the BGPS source extraction software, Bolocat (Rosolowsky et
al. 2009).  Bolocat utilizes a seeded watershed method (also referred
to as marker-controlled segmentation; Soille 1999) which is comprised
of two main steps.  First, regions with emission above a
signal-to-noise ratio of 2.5 are identified and expanded through a
nearest-neighbor algorithm to include adjacent, lower signal-to-noise
pixels.  Second, each identified region is subdivided into further
substructures based on the contrast between local maxima.  Bolocat
then determines source properties from emission-weighted moments.  Two
sets of coordinates are determined for each source: peak and centroid
coordinates.  The peak coordinates correspond to the peak of the 1.1
mm emission and are determined from median smoothed maps.  The
centroid coordinates are given by the first emission-weighted moments.
The major and minor axis dispersions ($\sigma_{maj}$ and
$\sigma_{min}$) are given by the second moments.  The object radius
(R$_{obj}$) is calculated from
\begin{equation} R_{obj}=\eta [(\sigma_{maj}^2 - \sigma_{beam}^2)(\sigma_{min}^2 - \sigma_{beam}^2)]^{1/4},
\end{equation}
where $\sigma_{beam} = \Theta_{FWHM}/\sqrt{8 \ln 2}$ and $\eta$ is a
factor relating the axis dispersions to the true size of the source,
found to be 2.4.  The value $\eta=2.4$ is adopted as the median value
derived from measuring the observed axis dispersions compared with the
true radius for a variety of simulated emission profiles spanning a
range of density distributions, sizes relative to the beam and
signal-to-noise ratios.  
The ellipses in Figures \ref{group1image}$-$\ref{group3image} mark the
positions and sizes of the extracted BGPS sources.  White ellipses 
denote BGPS sources which do not contain any \ammonia\ pointings or only 
contain a pointing which was not detected, and
black ellipses denote BGPS sources which contain at least one detected
\ammonia\ pointing.

\begin{deluxetable}{rc}
\tabletypesize{\scriptsize}
\tablewidth{0pt}
\tablecaption{\label{apercorrection}Aperture Flux Corrections}
\tablehead{
\colhead{Aperture} & \colhead{ } \\
\colhead{($^{\prime\prime}$)} & \colhead{\Snu(int)/\Snu(ap)} 
}
\startdata
40 & 1.46 \\
80 & 1.04 \\
120 & 1.01 
\enddata \\
\end{deluxetable}

Source flux densities are determined by
aperture photometry within standard apertures, 40\arcsec, 80\arcsec,
120\arcsec, and $2R_{obj}$ diameter apertures denoted as \Snu(40\as), 
\Snu(80\as), \Snu(120\as), and \Snu(obj).  An integrated flux density, 
\Snu(int), is also reported
for each source and is the sum of all pixels assigned
to a given source by the source extraction algorithm divided by the number
of pixels per beam.  The intensity in the BGPS maps was calibrated based on the
observed properties of planets.  Aguirre et al.~(2010) have compared the
BGPS data to other mm continuum surveys and found that we must
multiply our flux densities by 1.5$\pm$0.15 in order for our flux
densities to agree with the other surveys (see Aguirre et al. (2010) for 
more details).  In this paper we apply the flux calibration 
correction of 1.5$\pm$0.15 to the v1.0 BGPS products presented in Rosolowsky
et al. (2010) and released on the IPAC website.  All data products
and properties presented in this paper include this flux calibration factor.

Aperture corrections must be applied to the photometry in order to account
for power at large radii due to the non-gaussian side lobes of the 
CSO beam.  The aperture corrections were determined by 
comparing the integrated flux density for known point sources to the calculated
aperture flux densities.  The aperture correction values are listed in Table
\ref{apercorrection}. 
As can be seen from the aperture corrections, the 40\as\ aperture 
underestimates the flux density by almost 50\%.  The 80\as\
aperture only underestimates the flux density by approximately 4\%, 
and the 120\as\ aperture requires only a 1\%\ correction. 
The apertures corresponding to $2R_{obj}$ require varying aperture
corrections, while \Snu(int) represents all flux in the 1.1 mm maps
down to the 1$\sigma$ level and does not require an aperture 
correction factor.
The aperture corrections were determined based on sources very small
compared to the BGPS beam and therefore will slightly overestimate 
the flux densities of sources which are extended compared to the beam size.
For more details regarding source extraction and parameter estimation,
see Rosolowsky et al. (2010).  

\subsection{\ammonia\ Parameter Estimation}\label{nh3parameterestimation}
We estimate physical properties of the molecular gas by fitting a
simple model to each calibrated spectrum, following the method
described in Rosolowsky et al. (2008; hereafter R08).  The model
estimates the ammonia emission that would be observed from a slab of
molecular gas with a fixed optical depth in the NH$_3$(1,1) line of
$\tau$, LSR velocity $v_{\mathrm{LSR}}$, velocity dispersion
$\sigma_v$, radiative excitation temperature $T_{ex}$, and kinetic
temperature $T_{K}$.  For clarity, we note that $\tau$ here represents
the total opacity in the (1,1) transition which is then apportioned
into the 18 hyperfine components of the line according to their
statistical weights.  The central cluster of ammonia hyperfine
components, which is commonly studied in isolation, contains half the
total optical depth, $\tau_{main}=0.5\tau$.

The model spectrum provides a simultaneous estimate of all the ammonia
parameters; however, specific aspects of the spectral data control
each derived quantity.  For example, the ratio of strengths of the
(1,1) and (2,2) lines yields an estimate of the kinetic temperature.
The model accounts for the line widths by assuming a Gaussian velocity
distribution in the underlying slab of material so that the model
spectrum accounts for line broadening by hyperfine structure and
optical depth effects.  The difference between the line strength on
the $T_{mb}$ scale and the kinetic temperature provides information on
the radiative excitation of the line $T_{ex}$ under the assumption
that the emission fills the telescope beam (see \S\ref{excitationtemp}
for more details).  As noted in R08, the parameters $T_{ex}$ and
$\tau$ are strongly covariant and cannot be determined separately in
the low optical depth limit.  In these cases, only the product $\tau
T_{ex}$ is determined.  At higher optical depths, the relative
strengths of the central and satellite hyperfine transitions in the
(1,1) line provide an independent estimate of the optical depth.

Using these parameters as a description of the underlying molecular
gas, a model spectrum can be generated that adequately describes
emission we observe from the Gem OB1 sources including the hyperfine
structure of the ammonia transitions.  The best-fit set of parameters
is determined by a least-squares optimization of the model spectrum to
fit to all observed ammonia transitions simultaneously.  The primary
advantage of this approach is the straight-forward
estimation of errors and covariances in the fit parameters.  

\begin{figure}
\includegraphics[angle=0,width=3.4in]{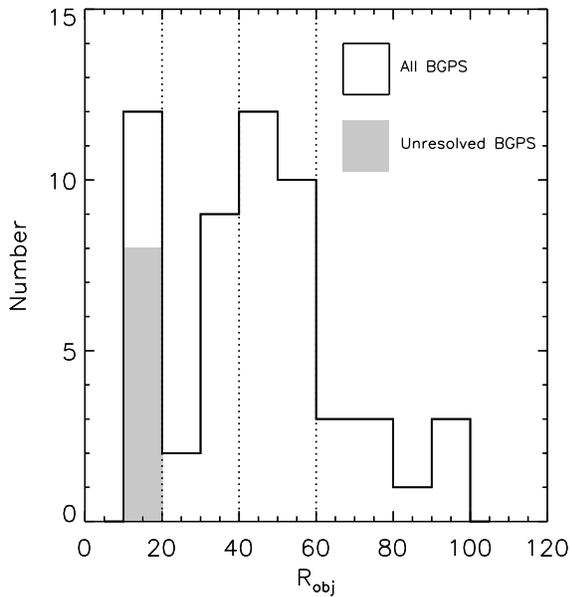}\caption{\label{histsize}
 The distribution of extracted source radii for all 55 BGPS sources. 
 The white histogram marks all 55 sources while the gray histogram
 marks the unresolved sources for which we have set the radius to an
 upper-limit of the beam size.  The dotted lines mark the radii
 of the aperture sizes used in the BGPS catalog, $R_{ap} =$ 20\as,
 40\as, and 60\as. }
\end{figure}

\begin{figure}
\includegraphics[angle=0,width=3.4in]{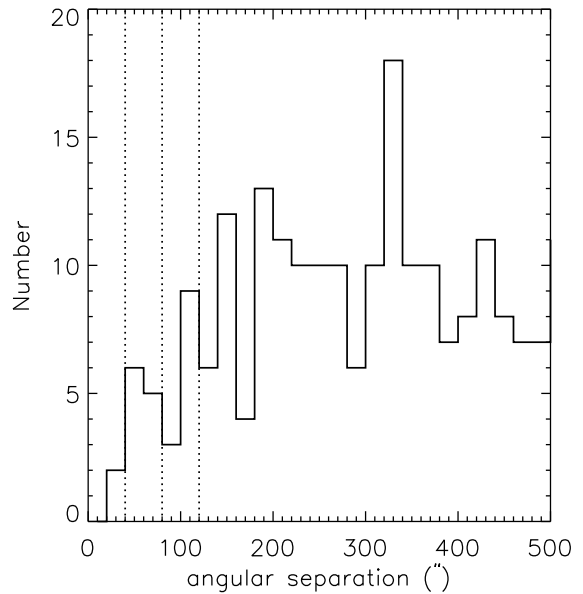}\caption{\label{histnearestneighbor}
 The distribution of distances between the 55 BGPS sources out to
 500\as.  The dotted lines mark the diameters of the apertures used
 in the BGPS catalog, $\theta_{ap} =$ 40\as, 80\as, and 120\as.  
 Only approximately 5 sources will suffer from contamination with
 the 120\as\ aperture.}
\end{figure}

\begin{deluxetable*}{rcccl}
\tabletypesize{\scriptsize}
\tablewidth{0pt}
\tablecaption{\label{mm_sources}Millimeter Sources}
\tablehead{
\colhead{ID} & \colhead{} & \colhead{BGPS ID} & \colhead{Group} & \colhead{}  \\                    
\colhead{Number} & \colhead{Source} & \colhead{Number} & \colhead{Number} & \colhead{Comments}    
}
\startdata
01........ & G189.646+00.131  & 7469  & 2 &  \\
02........ &  G189.659+00.185 & 7470 & 2 & \\
03........ & G189.682+00.185 & 7471  & 2 & IRAS 06047+2040 \\  
04........ & G189.782+00.265 & 7475  & 2 & IRAS 06053+2036 \\
05........ & G189.116+00.643 & 7467  & 1 & IRAS 06052+2122 \\
06........ &  G189.788+00.281 & 7479  & 2 &  \\
07........ & G189.713+00.335 &  7472 & 2 & \\ 
08........ & G189.789+00.291 & 7480 & 2 & \\
09........ & G189.744+00.335 & 7473 &  2 & \\
10........ & G189.782+00.323 & 7476 & 2 & \\
11........ & G189.836+00.303 & 7485 & 2 & IRAS 06055+2034\\
12........ & G189.950+00.231 & 7491  & 2 &  \\  
13........ & G189.776+00.343 & 7474  & 2 & S252A,  IRAS 06055+2039\\
14........ & G189.834+00.317 & 7484 & 2 & \\
15........ & G189.888+00.303 & 7489 & 2 & IRAS 06056+2032 \\
16........ & G189.804+00.355 & 7481 & 2 &  \\
17........ & G189.030+00.781 & 7465  & 1 & IRAS 06056+2131 \\
18........ & G189.831+00.343 & 7843  & 2 & J06084309+2036182\\  
19........ & G189.879+00.319 &  7487 & 2 & \\   
20........ & G189.885+00.319 & 7488  & 2 & \\ 
21........ & G189.810+00.369 & 7482  & 2 & \\
22........ & G189.032+00.793 & 7466  & 1 & IRAS 06056+2131\\
23........ & G189.921+00.331 & 7490 & 2 & \\
24........ & G189.015+00.823 & 7464  & 1 & \\   
25........ & G188.991+00.859 & 7463 &  1 & \\
26........ & G188.948+00.883 & 7461  & 1 & S252\\
27........ & G189.951+00.331 & 7492  & 2 &  \\   
28........ & G189.783+00.433 & 7477  & 2 & IRAS 06059+2042\\
29........ & G188.975+00.911 & 7462 & 1 & \\
30........ & G189.990+00.353 & 7493 & 2 & \\
31........ & G189.783+00.465 & 7478  & 2 & \\
32........ & G188.792+01.027 & 7460  & 1 & IRAS 06061+2151 \\
33........ & G190.006+00.361 & 7494 & 2 & IRAS 06061+2028 \\
34........ & G189.864+00.499 & 7486  & 2 & IRAS 06063+2040 \\
35........ & G189.231+00.893 & 7468  & 1 & IRAS 06065+2124  \\
36........ & G190.054+00.533 & 7496 &  2 & IRAS 06068+2030 \\
37........ & G190.044+00.543 & 7495 & 2 & \\
38........ & G190.063+00.679 & 7497 &  2 & \\
39........ & G190.192+00.719 & 7499 & 2 & \\
40........ & G190.171+00.733 & 7498  & 2 & IRAS 06078+2030  \\
41........ & G190.240+00.911 & 7500 & 2 & IRAS 06087+2031 \\
42........ & G192.629-00.157 & 7505  & 3 & \\
43........ & G192.602-00.143 & 7503 & 3 & S256, IRAS 06096+1757 \\
44........ & G192.629-00.123 & 7504 & 3 & \\
45........ & G192.581-00.043 & 7501  & 3 & S255N\\
46........ & G192.662-00.083 & 7507 & 3 & \\
47........ & G192.596-00.051 & 7502  & 3 & S255IR,  J06125330+1759215\\  
48........ & G192.644+00.003 & 7506  & 3 & S258\\
49........ & G192.719+00.043 & 7508  & 3 & IRAS 06105+1756 \\   
50........ & G192.764+00.101 & 7509  & 3 & \\
51........ & G192.816+00.127 & 7510 & 3 & \\
52........ & G192.968+00.093 & 7511  & 3 & \\
53........ & G193.006+00.115 & 7514 & 3 & \\
54........ & G192.981+00.149 & 7512  & 3 & J06142359+1744469\\
55........ & G192.985+00.177 & 7513 & 3 & 
\enddata
\end{deluxetable*}

The spectral model has two significant differences from that of the
R08 study.  In particular, the inclusion of NH$_3$(3,3) lines in the
analysis necessitates treating the ortho- and para-NH$_3$ states in
more detail.  Rather than simply adopting a full thermal equilibrium
as done by R08, we treat the two sets of states as different species
with an ortho-to-para ratio of 1:1, which is the high temperature
limit ($T_K>40$~K; Takano et al. 2002).  Given that the kinetic
temperature derived from only the (1,1) and (2,2) lines provides a
good estimate of the (3,3) line in most cases, this assumption is
adequate.  As such, partition functions for the two species are
decoupled and contain only the ortho- and the para- states:
\begin{align}
Z_O=& \notag\\
1+&\sum_{J}2(2J+1)\exp\left\{\frac{-h[BJ(J+1)+(C-B)J^2]}{kT_k}\right\} \notag\\ 
\qquad &\qquad \mbox{for } J=3,6,9,\dots  \label{partitionfunctionO} 
\end{align}
\begin{align}
Z_P=&\notag\\
\sum_{J}&(2J+1)\exp\left\{\frac{-h[BJ(J+1)+(C-B)J^2]}{kT_k}\right\} \notag\\
\qquad & \qquad \mbox{for } J=1,2,4,5,\dots,\label{partitionfunctionP}
\end{align}  
where we have only included the metastable states given by $J=K$.  The
lifetimes of the non-metastable states are relatively short and the
levels are not well populated.  The partition functions are subject to
the constraint that $Z_O=Z_P$, representing our assumption of the 1:1
ortho-to-para ratios.  In addition to a different partition function,
we also model the effects of the frequency switching explicitly in the
model spectrum since the frequency switch is smaller than in the R08
study.  We build a frequency switched version of the spectrum by
shifting it $\pm5$ MHz in frequency and subtracting half of the
amplitude in the model.  Thus, we fit both the peaks and the negative
features in the observed spectra, resulting in a fit which is more
sensitive to the low-intensity emission.

Emission from the H$_2$O and C$_2$S lines is not treated in detail by
the model but rather by simply calculating its velocity-integrated
intensity.  Since the model is necessarily simple, complicated sources
will be represented by average properties.  However, the model
produces good fits for the wide variety of sources found in the
observations.

\section{Results}\label{results}

\begin{deluxetable*}{llccccccccrr}
\tabletypesize{\scriptsize}
\tablewidth{0pt}
\tablecaption{\label{mm_props}Observed 1.1 mm Properties}
\tablehead{
\colhead{ID} & \colhead{} & \colhead{RA} & \colhead{Dec} & \colhead{peak RA} & \colhead{peak Dec} & \colhead{R$_{major}$} & \colhead{R$_{minor}$} & \colhead{PA} & \colhead{R$_{obj}$\tablenotemark{a}} & \colhead{S$_{\nu}$(120\as)} & \colhead{S$_{\nu}$(int)} \\
\colhead{number} & \colhead{Source} & \colhead{(J2000)} & \colhead{(J2000)} & \colhead{(J2000)} & \colhead{(J2000)} & \colhead{(\as)} & \colhead{(\as)} & \colhead{(\degree)} & \colhead{(pc)} & \colhead{(mJy)} & \colhead{(mJy)}
}
\startdata
01........ & G189.646+00.131 & 06 07 30.6 &  20 40 00.6 & 06 07 30.9 &  20 39 46.1 &  33.0 &  19.9 &  74 &  0.50 &     936 (250) &     988 (120) \\
02........ & G189.659+00.185 & 06 07 44.4 &  20 40 49.6 & 06 07 44.7 &  20 40 36.6 &  20.5 &  14.1 & 139 &  $<0.17$ & 1010 (270) &     370 (91) \\
03........ & G189.682+00.185 & 06 07 46.9 &  20 39 39.6 & 06 07 47.5 &  20 39 27.4 &  33.5 &  19.2 & 138 &  0.49 &    1509 (300) &    1700 (150) \\
04........ & G189.782+00.265 & 06 08 17.3 &  20 36 37.9 & 06 08 17.9 &  20 36 32.6 &  28.2 &  18.0 & 145 &  0.40 &    1380 (300) &    1190 (140) \\
05........ & G189.116+00.643 & 06 08 19.9 &  21 22 26.3 & 06 08 19.8 &  21 22 29.2 &  31.2 &  19.9 &   4 &  0.49 &    1190 (320) &    1090 (130) \\
06........ & G189.788+00.281 & 06 08 22.8 &  20 36 38.0 & 06 08 22.2 &  20 36 41.7 &  21.7 &  17.6 &  43 &  0.32 &    1630 (330) &    1019 (140) \\
07........ & G189.713+00.335 & 06 08 25.4 &  20 42 04.8 & 06 08 25.1 &  20 42 08.9 &  26.3 &  20.6 & 158 &  0.45 &    1360 (300) &    1160 (140) \\
08........ & G189.789+00.291 & 06 08 25.8 &  20 37 04.9 & 06 08 24.7 &  20 36 52.8 &  16.5 &   9.1 &  13 &  $<0.17$ & 1430 (330) &     347 (94) \\
09........ & G189.744+00.335 & 06 08 29.0 &  20 41 04.7 & 06 08 28.8 &  20 40 34.5 &  28.6 &  20.9 &  72 &  0.48 &    2960 (420) &    1960 (210) \\
10........ & G189.782+00.323 & 06 08 29.3 &  20 38 13.1 & 06 08 30.9 &  20 38 13.9 &  28.5 &  22.7 &  38 &  0.51 &    3190 (450) &    2240 (220) \\
11........ & G189.836+00.303 & 06 08 30.5 &  20 34 33.3 & 06 08 33.1 &  20 34 49.0 &  34.2 &  23.4 & 142 &  0.59 &    2480 (380) &    1780 (210) \\
12........ & G189.950+00.231 & 06 08 31.0 &  20 26 49.9 & 06 08 31.1 &  20 26 44.4 &  34.4 &  22.6 &  17 &  0.58 &    1670 (310) &    1910 (160) \\
13........ & G189.776+00.343 & 06 08 34.4 &  20 39 18.2 & 06 08 34.6 &  20 39 07.7 &  38.5 &  26.8 & 151 &  0.70 &    8930 (850) &   11100 (410) \\
14........ & G189.834+00.317 & 06 08 35.0 &  20 35 10.5 & 06 08 36.0 &  20 35 19.7 &  23.4 &  17.1 &  32 &  0.33 &    2730 (400) &    1530 (170) \\
15........ & G189.888+00.303 & 06 08 38.9 &  20 32 06.5 & 06 08 39.6 &  20 32 05.3 &  20.6 &  13.5 & 165 &  $<0.17$ & 1370 (290) &     537 (110) \\
16........ & G189.804+00.355 & 06 08 39.7 &  20 38 02.0 & 06 08 40.8 &  20 38 00.5 &  23.4 &  22.2 &  40 &  0.44 &    8060 (790) &    4700 (320) \\
17........ & G189.030+00.781 & 06 08 40.2 &  21 31 00.0 & 06 08 40.1 &  21 31 00.8 &  26.9 &  22.8 &  85 &  0.50 &   11100 (990) &   11200 (410) \\
18........ & G189.831+00.343 & 06 08 41.3 &  20 36 04.8 & 06 08 41.6 &  20 36 11.4 &  36.2 &  28.9 &  73 &  0.71 &    5780 (610) &    6420 (350) \\
19........ & G189.879+00.319 & 06 08 43.2 &  20 33 49.9 & 06 08 42.2 &  20 32 58.4 &  39.2 &  21.9 &  58 &  0.61 &    2390 (350) &    2400 (210) \\
20........ & G189.885+00.319 & 06 08 44.5 &  20 32 12.0 & 06 08 42.9 &  20 32 39.5 &  32.5 &  22.7 &   1 &  0.56 &    2450 (360) &    2080 (200) \\
21........ & G189.810+00.369 & 06 08 44.6 &  20 38 18.5 & 06 08 44.7 &  20 38 06.1 &  54.4 &  31.8 &  66 &  0.95 &    8090 (780) &   11300 (480) \\
22........ & G189.032+00.793 & 06 08 45.7 &  21 31 51.7 & 06 08 43.1 &  21 31 15.4 &  30.9 &  23.7 &  39 &  0.56 &   11100 (1000) &    6690 (430) \\
23........ & G189.921+00.331 & 06 08 49.2 &  20 31 46.8 & 06 08 50.1 &  20 31 07.1 &  25.0 &  14.9 &  21 &  0.25 &    2500 (380) &     697 (160) \\
24........ & G189.015+00.823 & 06 08 49.7 &  21 33 49.2 & 06 08 47.9 &  21 32 58.1 &  34.9 &  16.2 &  67 &  0.39 &    2690 (390) &    1660 (190) \\
25........ & G188.991+00.859 & 06 08 53.1 &  21 35 32.0 & 06 08 53.0 &  21 35 16.5 &  30.2 &  12.6 &  51 &  $<0.17$ &  977 (270) &     707 (110) \\
26........ & G188.948+00.883 & 06 08 53.4 &  21 38 23.8 & 06 08 52.9 &  21 38 16.9 &  29.0 &  24.4 &  89 &  0.55 &   12899 (1200) &   14599 (450) \\
27........ & G189.951+00.331 & 06 08 53.6 &  20 29 37.3 & 06 08 53.8 &  20 29 32.6 &  53.5 &  34.9 & 157 &  0.99 &    5980 (620) &   10800 (390) \\
28........ & G189.783+00.433 & 06 08 55.5 &  20 41 16.0 & 06 08 55.8 &  20 41 19.6 &  43.6 &  22.4 &  44 &  0.66 &    1439 (300) &    1840 (160) \\
29........ & G188.975+00.911 & 06 09 03.6 &  21 38 03.8 & 06 09 02.7 &  21 37 37.5 &  26.6 &  16.2 &  56 &  0.33 &     718 (260) &     703 (95) \\
30........ & G189.990+00.353 & 06 09 03.7 &  20 28 18.0 & 06 09 03.4 &  20 28 11.3 &  22.3 &  14.0 &   3 &  $<0.17$ &  886 (280) &     499 (91) \\
31........ & G189.783+00.465 & 06 09 04.4 &  20 42 12.6 & 06 09 03.0 &  20 42 15.4 &  30.0 &  19.8 & 124 &  0.47 &     917 (270) &     949 (120) \\
32........ & G188.792+01.027 & 06 09 06.7 &  21 50 48.5 & 06 09 06.0 &  21 50 39.1 &  27.2 &  22.2 &  83 &  0.49 &    4240 (570) &    4650 (250) \\
33........ & G190.006+00.361 & 06 09 08.4 &  20 27 33.7 & 06 09 07.2 &  20 27 34.9 &  30.5 &  20.1 & 110 &  0.48 &     887 (260) &     760 (110) \\
34........ & G189.864+00.499 & 06 09 20.7 &  20 39 30.8 & 06 09 20.6 &  20 39 02.7 &  50.2 &  36.5 &  27 &  0.99 &    3850 (450) &    6969 (300) \\
35........ & G189.231+00.893 & 06 09 30.4 &  21 23 44.7 & 06 09 30.6 &  21 23 39.8 &  17.4 &  15.4 & 100 &  0.20 &     464 (260) &     427 (79) \\
36........ & G190.054+00.533 & 06 09 51.1 &  20 30 14.5 & 06 09 51.7 &  20 30 03.2 &  52.7 &  23.4 &  21 &  0.75 &    2220 (340) &    2930 (220) \\
37........ & G190.044+00.543 & 06 09 54.6 &  20 30 56.2 & 06 09 52.7 &  20 30 52.2 &  25.1 &  17.7 &  22 &  0.37 &    2420 (360) &    1019 (170) \\
38........ & G190.063+00.679 & 06 10 27.0 &  20 33 48.5 & 06 10 25.7 &  20 33 45.6 &  20.3 &  12.9 & 149 &  $<0.17$ &  418 (230) &     229 (64) \\
39........ & G190.192+00.719 & 06 10 50.3 &  20 27 51.6 & 06 10 50.5 &  20 28 11.4 &  16.4 &  12.1 &  29 &  $<0.17$ &  685 (260) &     316 (74) \\
40........ & G190.171+00.733 & 06 10 51.3 &  20 29 57.7 & 06 10 51.2 &  20 29 38.8 &  45.1 &  16.9 &  32 &  0.49 &    1540 (300) &    2090 (160) \\
41........ & G190.240+00.911 & 06 11 40.4 &  20 31 26.7 & 06 11 39.6 &  20 31 12.8 &  28.0 &  17.2 &  73 &  0.38 &     726 (260) &     626 (98) \\
42........ & G192.629-00.157 & 06 12 33.6 &  17 54 48.7 & 06 12 33.5 &  17 54 40.4 &  21.9 &  16.7 &  57 &  0.30 &     796 (290) &     558 (96) \\
43........ & G192.602-00.143 & 06 12 33.6 &  17 56 19.8 & 06 12 33.2 &  17 56 33.1 &  25.9 &  16.8 &  11 &  0.35 &     894 (300) &     868 (110) \\
44........ & G192.629-00.123 & 06 12 39.8 &  17 55 28.8 & 06 12 41.0 &  17 55 39.2 &  35.6 &  21.3 &  16 &  0.56 &     709 (290) &     740 (110) \\
45........ & G192.581-00.043 & 06 12 52.8 &  18 00 32.6 & 06 12 52.9 &  18 00 29.0 &  25.2 &  21.9 & 174 &  0.46 &   14600 (1300) &   11400 (450) \\
46........ & G192.662-00.083 & 06 12 53.1 &  17 55 21.0 & 06 12 53.7 &  17 55 07.3 &  33.5 &  11.7 &  43 &  $<0.17$ &  565 (310) &     463 (85) \\
47........ & G192.596-00.051 & 06 12 54.1 &  17 58 54.1 & 06 12 52.8 &  17 59 31.0 &  46.5 &  29.5 &  22 &  0.83 &   15400 (1400) &   17300 (590) \\
48........ & G192.644+00.003 & 06 13 09.8 &  17 58 40.3 & 06 13 10.6 &  17 58 32.6 &  28.1 &  22.8 &  45 &  0.51 &    1330 (320) &    1380 (140) \\
49........ & G192.719+00.043 & 06 13 28.0 &  17 56 02.1 & 06 13 28.6 &  17 55 41.6 &  43.2 &  27.3 & 120 &  0.76 &    2320 (390) &    3580 (220) \\
50........ & G192.764+00.101 & 06 13 46.7 &  17 55 00.4 & 06 13 46.8 &  17 55 02.5 &  39.6 &  21.0 & 146 &  0.59 &    2020 (370) &    2480 (180) \\
51........ & G192.816+00.127 & 06 13 58.8 &  17 52 51.2 & 06 13 58.8 &  17 53 03.0 &  20.1 &  13.2 &  39 &  $<0.17$ &  615 (320) &     505 (84) \\
52........ & G192.968+00.093 & 06 14 08.3 &  17 44 04.6 & 06 14 09.6 &  17 44 03.9 &  25.4 &  16.2 & 166 &  0.32 &     682 (310) &     620 (93) \\
53........ & G193.006+00.115 & 06 14 18.8 &  17 42 36.9 & 06 14 19.1 &  17 42 41.7 &  19.4 &  14.8 & 162 &  0.19 &    1310 (350) &     661 (110) \\
54........ & G192.981+00.149 & 06 14 23.4 &  17 44 28.4 & 06 14 23.7 &  17 44 56.1 &  49.6 &  25.5 &  43 &  0.78 &    4290 (520) &    6480 (290) \\
55........ & G192.985+00.177 & 06 14 30.2 &  17 45 34.0 & 06 14 30.4 &  17 45 31.6 &  18.4 &  14.3 &  85 &  $<0.17$ &  844 (320) &     516 (90) 
\enddata 
\tablecomments{Errors are given in parentheses.}
\tablenotetext{a}{Sources that are unresolved with the BGPS beam are assigned the beam size as an upper-limit on the object radius.}
\end{deluxetable*}

\subsection{1.1 mm Results}
We have extracted 55 millimeter continuum sources from the BGPS maps. 
The 1.1 mm sources are listed in Table
\ref{mm_sources}, along with their ID number in the complete BGPS
source catalog, their corresponding group number (see
\S\ref{mmimagesandcatalog}), and any coincident IRAS sources, 2MASS
sources and well known millimeter sources.  
Table \ref{mm_props} lists extracted 1.1 mm source properties including 
a running source ID number (column 1), the source name (based on
the Galactic coordinates of the peak emission; Column 2), centroid RA
and DEC (columns 3 and 4), RA and DEC of the peak emission (columns 5
and 6), size of semi-major and semi-minor axes in arcseconds (columns
7 and 8), position angle of best-fit ellipse measured
counter-clockwise from the $x$-axis ($\ell=0^{\circ}$) in Galactic
coordinates (column 9), the deconvolved source radius $R_{obj}$
(column 10), and \Snu(120\as) and \Snu(int) (columns 11 and 12).  
Some sources are
sufficiently small that a deconvolved radius cannot be determined, and we
assume a source diameter equal to the beam FWHM as an upper limit on
the radius for these objects.  

\Snu(int) and $R_{obj}$ adequately characterize the
sources but these two quantities are highly 
dependent upon the source extraction algorithm and are not easily 
compared with other observations.  In order to support comparison 
of our results with other studies, we will also present a flux 
within a well-characterized and commonly used aperture.  

\stepcounter{figure}
\begin{figure*}
\begin{center}
\figurenum{\arabic{figure}a}
\end{center}
\epsscale{0.8}
\plotone{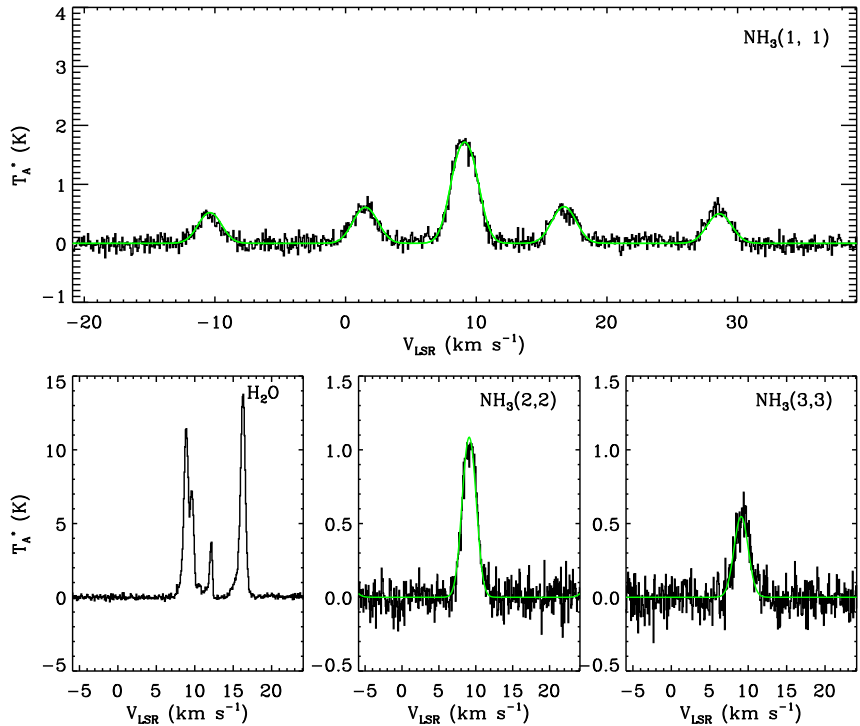} \figcaption{\label{nh3spectrum} 
H$_2$O,\ammonia(1,1), (2,2), and (3,3) spectra for G189.776+0.343 (ID \#13).
Top panel: Black is observed \ammonia(1,1) spectrum and green is the
model fit.  From left to right, the bottom panels show H$_2$O,
\ammonia(2,2) and \ammonia(3,3) spectra.  Figures 8b-at are available in the 
 online version of the Journal.}
\end{figure*}

To evaluate which aperture most closely represents the true mm flux,
we consider the derived object radii as well as the distance between
all BGPS sources.  
Figure \ref{histsize} shows the distribution of derived source radii for 
the 55 BGPS sources (white histogram) and the unresolved sources
where we have assigned the beam size as an upper-limit for the source
radius (gray histogram).  The dotted lines mark the radii of the
apertures in which we present photometry in the BGPS catalog.  
There is a large spread
in the object radii, with the largest being almost 100\as.  
Figure \ref{histnearestneighbor} shows the distribution of angular separation 
between the emission maxima of all 55 BGPS sources out to 500\as.  
The dotted lines mark the diameters of our chosen apertures.  

To characterize the mm flux well, we should choose an aperture size
greater than the largest source radius yet small enough to reduce
the number of sources which will include flux from their neighbors.
If we choose an aperture much larger than the source and
no other mm sources fall within that aperture, then the flux
will still be representative of the total mm flux since the background
averages to zero.  The distribution of source sizes supports the
use of the 120\as\ aperture, which is only smaller than 10 of the 55
BGPS sources.  Figure \ref{histnearestneighbor} shows that for the 
120\as\ aperture, only approximately 5 source fluxes will be 
contaminated by emission from nearby sources.

We have chosen to present the \Snu(120\as) fluxes as well-characterized
fluxes for comparison with other studies.  We present masses calculated
from both \Snu(120\as) and \Snu(int).  However, when
calculating physical properties which are highly dependent upon 
size, such as the volume and surface densities, we will
present \Snu(int) and use the derived source radius as the physical
size.  

\begin{figure}[b]
\epsscale{1.0}
\plotone{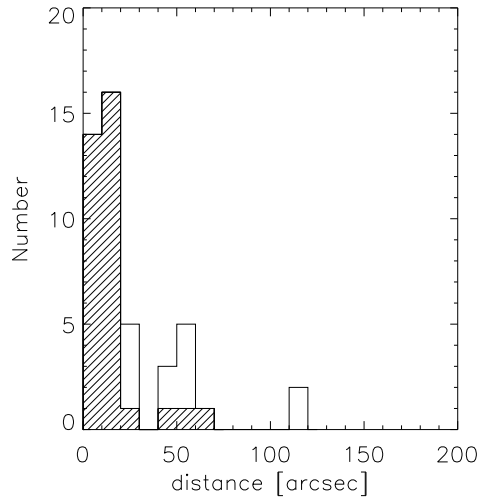}\figcaption{\label{bgps-nh3_dist}
 The white histogram denotes the distance between each of the 46
 cross-matched \ammonia\ pointings and the peak of the associated
 BGPS source.  The striped histogram marks the distance between the
 peak in 1.1 mm emission and the nearest \ammonia\ pointing for the
 34 BGPS sources in the full sample.  The \ammonia\ pointings were
 chosen by eye from preliminary BGPS images, which can account for
 the offsets larger than the beam size in the striped histogram.  The
 large distances in the white histogram result from multiple
 \ammonia\ pointings per BGPS source, with typical spacings between
 pointings on order of the GBT beam size (31\as) or larger.  }
\end{figure}

\begin{figure*}
\includegraphics[angle=0,width=3.4in]{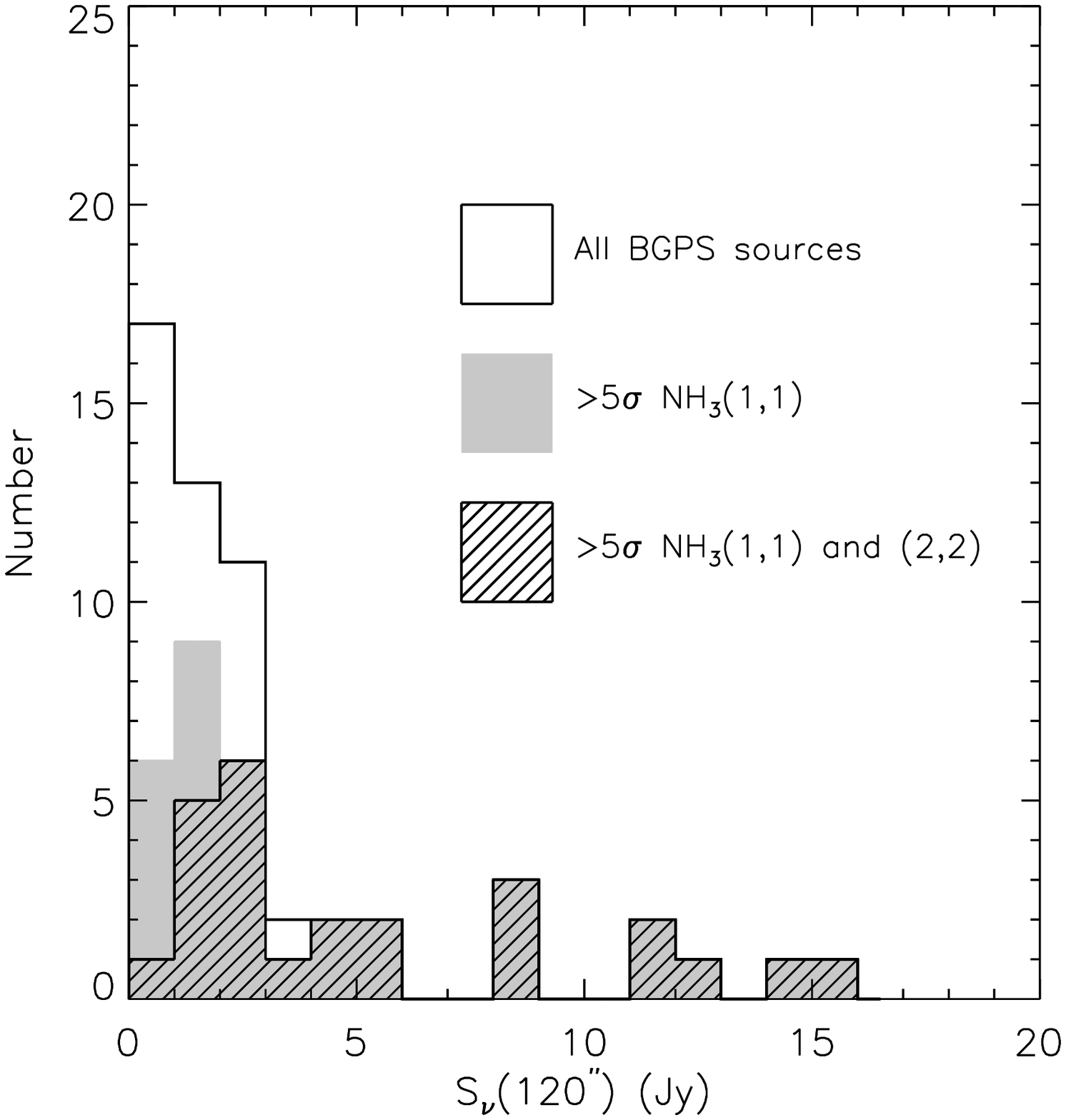}
\includegraphics[angle=0,width=3.4in]{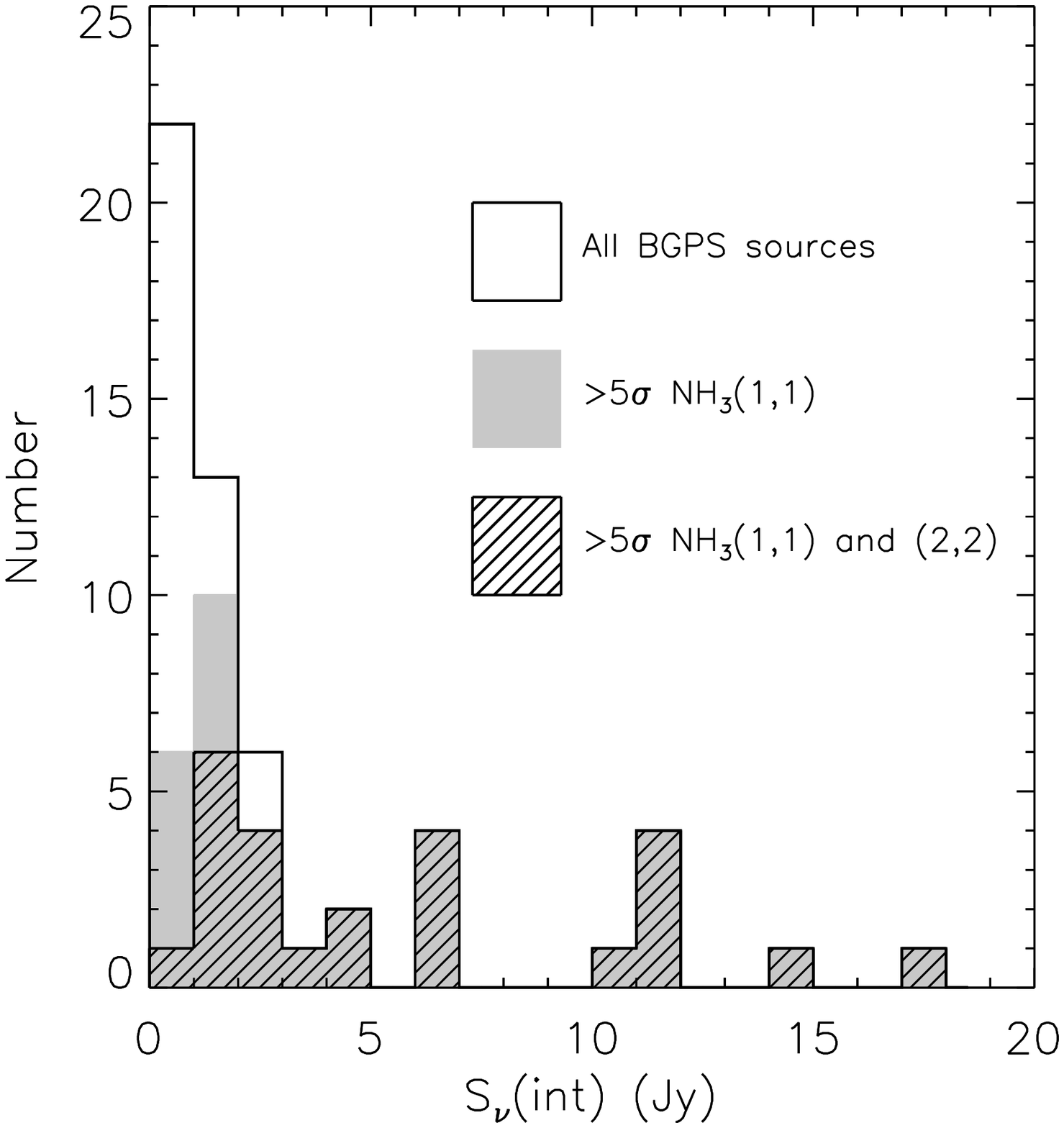} 
\caption{\label{fluxdists} 
Distribution of \Snu(120\as) (left) and integrated flux densities, \Snu(int) (right).  The white histograms include all extracted 1mm sources, the gray histograms represent the full sample (all sources with a $>5\sigma$ detection in \ammonia(1,1)), and the striped histograms represent subset 1 (all sources with a $>5\sigma$ detection in \ammonia(1,1) and (2,2)).  
}
\end{figure*}

\subsection{\ammonia\ Results}

Example spectra of the \ammonia(1,1), (2,2), (3,3) and H$_2$O lines
are shown in Figure \ref{nh3spectrum} for the source G189.776+00.343
(ID \#13).  Spectra for all \ammonia\ pointings are shown in 
Figures 8b$-$at in the online materials.  \ammonia(4,4) and C$_2$S 
were not observed for many sources
and are not shown here.  We detected a $> 5\sigma$ \ammonia(1,1) line
in 95\% (53 of 56) of all \ammonia\ pointings, and 60\% (33) of the
pointings had at least a 5$\sigma$ detection in both the (1,1) and
(2,2) lines, where a $5\sigma$ detection means the integrated intensity
is greater than 5 times the measured error in the integrated intensity
for each source.

\subsection{Combined 1.1 mm and \ammonia\ Results}

Each \ammonia\ pointing was assigned to a single
distinct BGPS source.  Of the 55 BPGS sources, 37 have ammonia observations 
along a
line of sight that intersects the source.  Three BGPS sources have
an ammonia pointing associated with the source that lacks any
discernible NH$_3$(1,1) emission.  In the remaining 34 BGPS sources,
there may be multiple \ammonia\ pointings within the contours of the
BGPS source, and the \ammonia\ pointings do not necessarily correspond
to either the centroid or peak coordinates of each BGPS source because
they were chosen by eye prior to completion of the final BGPS maps.
We exclude the seven \ammonia\ pointings that do not fall within the
contours of an extracted BGPS source from the analysis.  
These pointings correspond to
low significance BGPS sources.  The seven pointings all
have significant NH$_3$(1,1) emission, indicating that many marginal
detections in the BGPS not included in the catalog are likely real.

In this analysis we will consider only the \ammonia\ pointing which is
located closest to the peak of the millimeter emission.  
The angular separation between
the closest \ammonia\ pointing and the peak of the millimeter emission is
shown in Figure \ref{bgps-nh3_dist}.  The white histogram includes all
\ammonia\ pointings while the striped histogram includes only the
\ammonia\ pointings closest to the millimeter peak.  Most of the
\ammonia\ pointings assigned to a BGPS source are located less than a
FWHM beam away from the millimeter peak.

We define a ``full sample'' which consists of 34 BGPS sources that 
have at least a
5$\sigma$ detection in the \ammonia(1,1) transition.  Sources without
a 5$\sigma$ detection in the \ammonia(2,2) transition can only provide
an upper-limit to the derived gas properties such as \tk.
Consequently, the full sample can only characterize upper-limits for
these properties.  We also define ``subset 1'' which is comprised of 25
BGPS sources with a $>5\sigma$ detection in \textbf{both} the
\ammonia(1,1) and (2,2) transitions.  However, subset 1 may not be
characteristic of the faint end of the full sample.  Due to the
complementary properties of each sample, we will present the
characteristics of both.  For clarity, we point out that subset 1 is
entirely contained in the full sample.

Observed properties of \ammonia\ pointings in the full sample are
given in Table \ref{nh3_props}, including the source ID (column 1),
specific coordinates of the \ammonia\ pointing which do not necessarily
correspond to the peak of the millimeter emission (Columns 2 and 3),
radial velocity and Gaussian $1/e$ line width (columns 4 and 5), 
and main beam temperature
and integrated intensities for the \ammonia(1,1), (2,2), (3,3) and (4,4) 
transitions (columns 6-13).  For undetected lines, we present upper-limits
for the main beam temperature and integrated intensities given as
\tmb$< 5$ RMS and $W<5$ RMS $\Delta v / \sqrt(N)$, where RMS is
the noise per channel of each spectrum, $\Delta v$ is the channel width 
in velocity, and $N$ is the number of channels used to calculate the RMS.

Table \ref{gas_props} lists derived gas properties including 
the measured optical depth of the (1,1) 
transition (column 2), kinetic gas temperature (column 3; see \S\ref{kinetictemp}),
excitation temperature of the \ammonia(1,1) transition (column 4), derived
sound speed and non-thermal velocity dispersion ($a$ and $\sigma_{NT}$; 
columns 5 and 6; see \S\ref{vel_disp}), and flags for detection or non-detection
of the H$_2$O and CCS lines (columns 7 and 8).  
The additional 12 \ammonia\ pointings are 
included in Tables \ref{nh3_props} and \ref{gas_props}, but are 
not included in the general analysis.  Numbers given in parentheses are errors 
in the same units as each column. 

The distributions of the flux densities, \Snu(120\as) and \Snu(int), are
shown in Figure \ref{fluxdists}.  The white histogram includes all 55
BGPS sources extracted from the millimeter maps.  The gray
histogram includes all BGPS sources included in the full sample
($>5\sigma$ detection in \ammonia(1,1)), while the striped histogram
denotes subset 1 ($>5\sigma$ detection in \ammonia(1,1) and (2,2)).
The faintest sources are excluded from subset 1, as expected.  

\begin{figure}
\includegraphics[angle=0,width=3.4in]{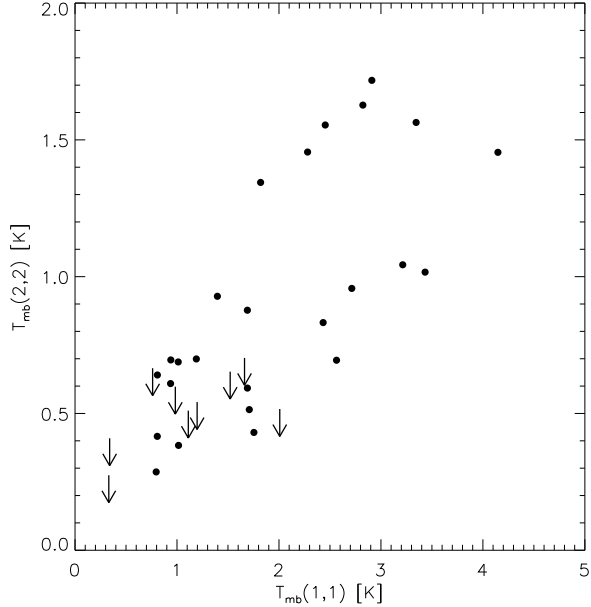} \caption{\label{tmbhist} 
Peak main beam temperature of the \ammonia(1,1) line versus the (2,2) line.  Sources in subset 1 are marked by solid circles.  Sources excluded from subset 1 give only upper limits for the (2,2) transition and are marked by arrows.  
}
\end{figure}

Figure \ref{tmbhist} plots the peak \tmb\ of
the (2,2) transition versus the peak \tmb\ of the (1,1) transition.
Pointings with $<5\sigma$ detection in the (2,2) transition are shown
as upper limits.  The fainter lines in (1,1) are typically the
pointings with a weak or non-detection of (2,2).  The weaker BGPS
sources typically have weaker \ammonia(1,1) detections and only
upper-limits for the \ammonia(2,2) line.

\begin{figure*}
\epsscale{1.0}
\plotone{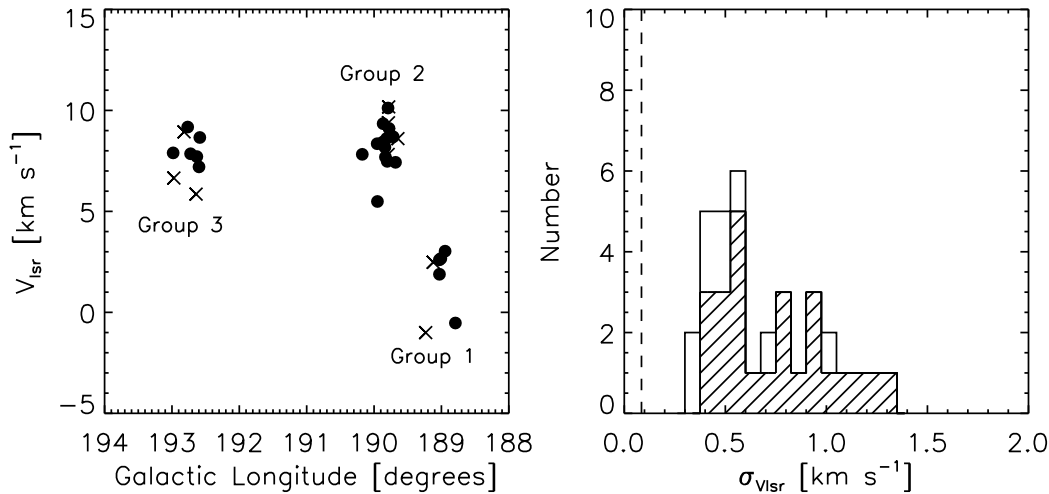} \figcaption{\label{vlsr_sigmav} 
Left:  Velocity with respect to the LSR versus Galactic Longitude.  Groups 2 and 3 span the same velocity range, while Group 1 is found at lower velocities.  Right:  Distribution of velocity dispersions.  The white histogram marks the full sample and the dashed histogram marks subset 1.  The dashed line marks the spectral resolution
of the \ammonia\ spectra, 0.08 \kms.  
}
\end{figure*}

\subsection{Maser Activity}

We also obtained observations of the 22 GHz water maser in conjunction 
with our GBT \ammonia\ observations.
Eight of the 34 BGPS sources presented here have water maser emission
(see columns 7 and 8 of Table \ref{gas_props}).
Of these, only Sources \#16 and \#24 have not been reported previously
as maser sources.  The remaining 6 BGPS sources with maser activity
were previously known to exhibit water as well as some SiO or methanol
maser activity in a $1'$ region around the observed GBT position based
on a SIMBAD search.  The fraction, $8/34=0.2$, can be regarded as a
lower limit as two sources, \#13 and \#47, show maser emission along one
line of sight but not another line of sight through the same core.
Since the GBT ammonia spectroscopy was not complete mapping, some
sources may have maser emission elsewhere in the object.  However, the
masing lines of sight are those nearest the peaks of the BGPS
emission, so the lower limit is likely representative of the true
fraction.  For comparison, Szymczak et al.~(2005) find a detection
fraction of water masers of 0.52 using 6.7 Ghz methanol masers,
associated with massive young stellar objects, as a prior.  Similarly,
Chambers et al.~(2009) find detection rates of 0.16, 0.54 and 0.59 for
water masers in their IRDC cores classified with Spitzer data as
``quiescent,'' ``red'' and ``active'' respectively.  The BGPS-selected
water maser detection fraction is most similar to the detection fraction 
in ``quiescent'' IRDCs determined by Chambers et al.~(2009). 
We see no evidence for maser activity in the NH$_3$(3,3)
transition. All (3,3) spectra are consistent with the temperatures and
line widths determined in conjunction with the (1,1) and (2,2) states.
We defer further discussion of the water maser detection fraction to
a later paper (M.~K.~Dunham et al., in prep).

\section{Analysis}\label{analysis}

\subsection{Radial Velocities and Distances\label{vlsr_and_d}}
We adopt the VLBI parallax distance to S252 of $2.10\pm0.026$ kpc
(Reid et al.~2009) for the entire sample of BGPS sources.  
The radial velocities observed
in the \ammonia\ spectra are listed in Table \ref{nh3_props} (column 4), 
and are
shown in Figure \ref{vlsr_sigmav}.  Filled circles mark sources in
Subset 1, while the crosses mark sources without a 5$\sigma$ detection
in the \ammonia(2,2) line.  The sources excluded from Subset 1 are
spread evenly in velocity space, as expected since the weaker BGPS
sources should not be preferentially located at any given radial
velocity.  Toward the
Galactic anti-center (as well as all Cardinal points: $l=0^{\circ},
90^{\circ}, 180^{\circ},$ and $270^{\circ}$), radial velocities
expected from purely circular rotation are 0 \kms, rendering kinematic
distances unsound.  A non-zero radial velocity along these lines of
sight is most likely due to the peculiar motions of individual gas
clouds rather than galactic rotation.  

From Figure \ref{vlsr_sigmav}, it is apparent that groups 2 and 3
span the same velocity range (5 to 10 \kms) while group 1 is found at
slightly lower radial velocities ($-$1 to 4 \kms).  Although there is
no evidence in our data that these groups are related, surveys of
lower density tracers (such as CO and $^{13}$CO) toward S247 and S252
(groups 1 and 2) show evidence of interaction between the two HII
regions.  K\"ompe et al. (1989) mapped the S247/252 complex in
$^{13}$CO ($J=1-0$) transition with a beam-sampled map at
4\am\ resolution and found a bridge connecting the two clouds.
Specifically, the bulk of the material towards S247 is found between
velocities of $-$1 \kms\ and 5 \kms\ with a tail extending to
velocities of up to 10 \kms, which overlaps the S252 region in
velocity space (see Figure 7 in K\"ompe et al. 1989).  Carpenter et
al. (1995a) mapped a larger region including the S247 and S252
complexes in $^{13}$CO($J=1-0$) and found three distinct filaments.
Two filaments with radial velocities of $\sim$2 \kms\ wrap around
S247, while the third filament, at a radial velocity of $\sim$10 \kms,
extends to the S252 region.  It is clear that the S247 and S252
regions (groups 1 and 2) are interacting, justifying our application
of the VLBI parallax distance to our entire sample regardless of the
difference in radial velocities between the groups of sources.

Not all sources will be at precisely the parallax distance; however,
we account for variation in the distances by including a larger
error than the parallax distance error.  Since it is clear that groups
1 and 2 are part of filaments wrapping around and interacting between
S247 and S252, we will consider the size of the HII regions when
determining the spread in distance.  S247 has a diameter of 9\am\ and
S252 has a diameter of 40\am\ (Sharpless 1959).  If the cloud between
Groups 1 and 2 is assumed to be spherically symmetric, its angular
size on the sky can provide an estimate of the distance along the line
of sight.  Since it is apparent that groups 1 and 2 are interacting,
we will consider a sphere of approximately 100\am\ in diameter that
encompasses both groups 1 and 2 and the associated HII regions, and
corresponds to a distance uncertainty of 0.061 kpc.  We will assume
the spread in distance is approximately equal to the 100\am\ diameter,
and will assign that to the distance error to account for the case
where the parallax source is found at the extreme edge of the region
and note that this is likely a minimum estimate of the error.

\subsection{Kinetic Temperature\label{kinetictemp}}
The kinetic temperatures derived from the \ammonia\ observations are
given in column 3 of Table \ref{gas_props}.  The uncertainties in individual
determinations of \tk\ are small, with $\mean{\sigma_{\tk}/\tk} =
0.04$.  In
this and subsequent analyses, the uncertainties quoted are the
standard deviation about the mean of the sample, rather than the uncertainty in the
mean.  For sources without a 5$\sigma$ detection in
\ammonia(2,2) the derived kinetic temperature is an upper-limit, and
thus the mean kinetic temperature for the full sample is only an
upper-limit.  The distribution for the full sample is characterized by
$\langle$\tk$\rangle < 19.4\pm4.9$ K, while the distribution for
Subset 1 is characterized by $\langle$\tk$\rangle =20.6\pm 5.1$ K.  
The mean limit on kinetic temperature for the sources
without a 5$\sigma$ \ammonia(2,2) detection is $\langle$\tk$\rangle <
16.1\pm2.3$ K.  As some property distributions may be non-Gaussian, the
uncertainty should be regarded primarily as an indicator of the
scatter in the sample.  The dispersion about the mean kinetic temperature
for the full sample is thus six times
larger (0.24) than the fractional uncertainties in individual temperatures, 
so the differences among sources are real.  

Zinchenko et al.~(1997) mapped 17 molecular clouds associated with FIR sources
in \ammonia(1,1) and (2,2) and found that \tk\ increases toward the peak
of the \ammonia\ emission.  Additionally, Friesen et al.~(2009) have shown that
\ammonia\ emission closely follows mm continuum emission.  
Since our \ammonia\ pointings are located near the peak of the mm emission,
which corresponds to the \ammonia\ peak as well, the mean \tk\ is 
likely smaller than the peak \tk\ we report.

\subsection{Velocity Dispersion and Non-Thermal Pressures}\label{vel_disp}
The line-width provides a measure of the one-dimensional velocity
dispersion within each source.  Rosolowsky et al. (2008) found
velocity dispersions ranging from 0.07 to 0.7 \kms\ for the
\ammonia(1,1) and (2,2) lines of low-mass star-forming cores in the
Perseus molecular cloud, and they suggest that there is evidence of
multiple velocity components in all sources with $\sigma_v >$ 0.2
\kms.  The velocity dispersions are listed in column 5 of Table 
\ref{nh3_props} and the distribution of our sample is shown in
Figure \ref{vlsr_sigmav}.  In contrast to the low-mass sources in
Perseus, our distribution has a tail extending up to a high velocity
dispersion $\sigma_v \approx$ 1.3 \kms.

\begin{figure}
\epsscale{1.0}
\plotone{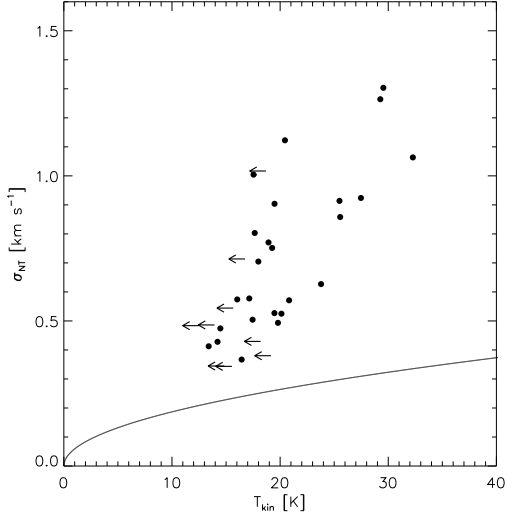}\figcaption{\label{magturb}
The non-thermal velocity dispersion, $\sigma_{NT}$, versus derived kinetic temperature.  Arrows indicate upper limits for the kinetic temperatures.  Solid line denotes the thermal sound speed given by $a=\sqrt{kT/\mu m_{H}}$.  
}
\end{figure}

The observed velocity dispersion is a combination of the thermal and
non-thermal motions of the gas.  In order to calculate the non-thermal
contribution, $\sigma_{NT}$, we simply remove the thermal
contribution:
\begin{equation}
\sigma_{NT}=\sqrt{\sigma_v^2-\frac{k\tk}{17 m_H}},
\label{sigmanonthermal}
\end{equation}
where $(k\tk/17 m_H)^{1/2}$ is the thermal broadening due to
\ammonia\ and $17 m_H$ is the mass of a single ammonia molecule.  The
non-thermal velocity dispersion ranges from 0.37 \kms\ to 1.3
\kms\ with a mean of $0.72\pm0.25$ \kms\ for subset 1.  We can compare
$\sigma_{NT}$ to the predicted thermal sound speed of the gas given by
$a=(k\tk/\mu m_H)^{1/2}$, where $\mu= 2.37$.
The thermal sound speed for subset 1 ranges from 0.22 \kms\ to 0.34
\kms\ with a mean of 0.27$\pm$0.03 \kms.  The mean Mach number is then
2.8.  Both $a$ and $\sigma_{NT}$ are listed in Table \ref{gas_props}
(columns 5 and 6).
The corresponding ratio of $\sigma_{NT}$ to $a$ ranges from 1.5 to 4.2
with a mean of 2.7$\pm$0.8.  The non-thermal velocity dispersion is
shown in Figure \ref{magturb} versus \tk.  The solid line plots the
thermal sound speed, $a$, as a function of \tk.  In all sources, the
non-thermal velocity dispersion is much larger than the thermal
velocity dispersion and the discrepancy increases with \tk.

\begin{figure*}
\epsscale{0.85}
\plotone{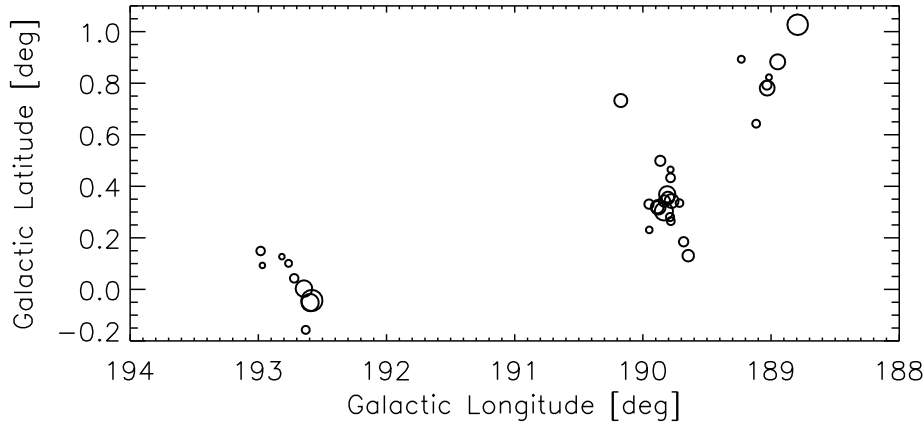}\figcaption{\label{glonglatmagturb}
Galactic longitude versus Galactic latitude for the full sample.  Symbol size denotes the magnitude
of the non-thermal velocity dispersion, $\sigma_{NT}$.  
}
\end{figure*}

There is no apparent correlation between $\sigma_{NT}$ and source size
or mass.  However, there is a correlation between the position of a
source within its group and $\sigma_{NT}$.  Figure
\ref{glonglatmagturb} displays the position of each source in Galactic
coordinates and the symbol size denotes the magnitude of
$\sigma_{NT}$.  For Group 2, the highest values of $\sigma_{NT}$ are
found at the center of the cluster.  A similar statement could be made for 
Group 3, but there it is difficult to definitively say, due to the smaller 
number of sources.

\subsection{Virial and Isothermal Masses\label{masses}}
Combining heterodyne and dust continuum observations provides
estimates of both virial mass and total gas and dust mass.  Total gas
and dust mass (also called the isothermal mass, \miso, due to the
simplifying assumption that the dust can be characterized by a single
temperature) is given by
\begin{align} M_{iso}&=\frac{S_{\nu}D^{2}}{\kappa_{\nu}B_{\nu}(T_d)}\notag\\
=13.1\ \mbox{M$_\odot$}\ \left(\frac{S_{\nu}}{1\ Jy}\right)&\left(\frac{D}{1\ kpc}\right)^2\left(\frac{e^{13.0\ K / T_d}-1}{e^{13.0\ K / 20.0\ K}-1}\right) ,
\label{miso}
\end{align}
where $S_{\nu}$ is flux density, $D$ is the distance, $\kappa_{\nu}$
is the dust opacity per gram of gas and dust, and $B_{\nu}$ is the
Planck function evaluated at a dust temperature, $T_d$.  We logarithmically
interpolate the Ossenkopf \& Henning dust opacities (1994, Table 1,
column 5, commonly referred to as OH5 dust) to the effective central
frequency of the Bolocam bandpass convolved with a 20 K blackbody
modified by an opacity varying with frequency as $\kappa_{\nu} \propto \nu^{1.8}$
(271.1 GHz; Aguirre et al. 2010) and obtain $\kappa_{\nu} = 0.0114$
cm$^2$ g$^{-1}$.

We assume the gas and dust are collisionally coupled, make the
simplifying assumption that $T_d=\tk$ and use \tk\ derived from the
\ammonia\ observations.  Evans, Blair, \& Beckwith (1977) observed the
S255 molecular cloud in \co, \coo, H$_2$CO, and 2.2 \um\ emission.
They studied the gas and dust energetics and concluded that the
primary energy flow is through infrared dust emission and that an
embedded source and the exciting stars of the many nearby HII regions
are the likely heat sources for the cloud.  Dust typically controls
the energetics in the dense interior of the star-forming regions where
the density is high enough to shield the gas from the ultraviolet
radiation in the interstellar radiation field (ISRF) and also to
transfer energy from the dust to the gas via collisions (Goldreich \&
Kwan 1974, Evans et al. 2001).  Towards the outer edges of the clump
the density is much lower, and the gas is no longer collisionally
coupled to the dust.  The gas is also now subject to the ultraviolet
portion of the ISRF, allowing the gas kinetic temperature to rise
above the dust temperature (see Figure 9 in Mueller et al. 2002).  The
BGPS observations probe all of these regions, and detailed models will
be needed to include temperature gradients. For now, we rely on
previous modeling (e.g., Mueller et al. 2002, Shirley et al. 2002) to
support the assumption that $T_d = \tk$ on average.

As discussed in \S\ref{kinetictemp}, our single measurement of \tk\
likely corresponds to the maximum temperature seen within the BGPS 
source.  Thus, using $T_d=\tk$ to describe the dust temperature for
the entire source will result in a lower-limit for the isothermal mass.  
Zinchenko et al.~(1997) find a maximum \tk\ of 28 K at
the peak of the \ammonia\ emission, and a minimum \tk\ of
15 K at the edge of the \ammonia\ emission.  This difference in
temperatures corresponds to a factor of 2.3 increase in mass if the 
minimum temperature is assumed to be $T_d$.  If we assume mean values
from Zinchenko et al.~(1997) of \tk=24 and 17.5 K, the mass would
increase by a factor of 1.5 assuming the minimum temperature as $T_d$
rather than the maximum.  If we are observing the maximum \tk\
then our calculated values of \miso\ are likely to underestimate 
the mass by a factor of up to approximately 2.  

The virial mass for a spherical, uniform density gas cloud is given by 
\begin{equation} M_{vir}=\frac{5\sigma_{v}^{2}R}{G}, \label{mvireqn}
\end{equation}
where $\sigma_{v}$ is the Gaussian 1/$e$ width of the line (model fit for the
velocity dispersion),
$R$ is the source radius, and $G$ is the gravitational constant.
We assume that the velocity dispersion measured with our single \ammonia\ pointing
is also applicable at the edge of the 1.1 mm continuum emission, and use
the single measured value in conjunction with the mm-derived object
radius, $R_{obj}$, to calculate the virial mass.  This assumption is
supported by Friesen et al. (2009) who
found that \ammonia\ emission closely traces 850 \um\ continuum emission
and covers the same spatial extent as the cold dust producing the 
continuum emission.  However, Zinchenko et al.~(1997) found that the 
ammonia line width increased toward the peak of the \ammonia\ emission
for about half of their sample.  Thus, we are likely overestimating the 
virial mass by using the line width measured near the peak.

\begin{figure*}
\plotone{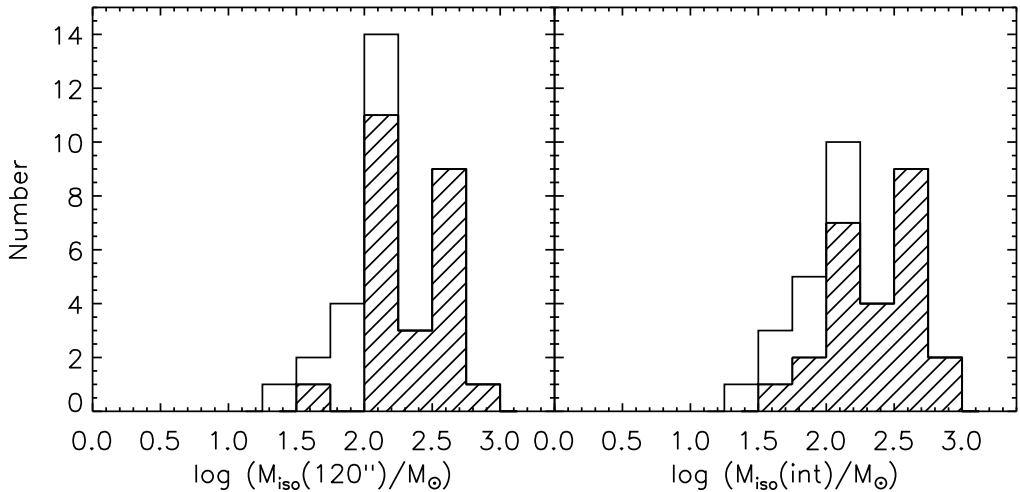} \figcaption{\label{masshists} 
Distributions of derived \miso\ derived from \Snu(120\as) (left) and \Snu(int)(right).  The white histogram marks the full sample and the striped histogram marks subset 1. 
}
\end{figure*}

\begin{figure}[t]
\epsscale{0.9}
\plotone{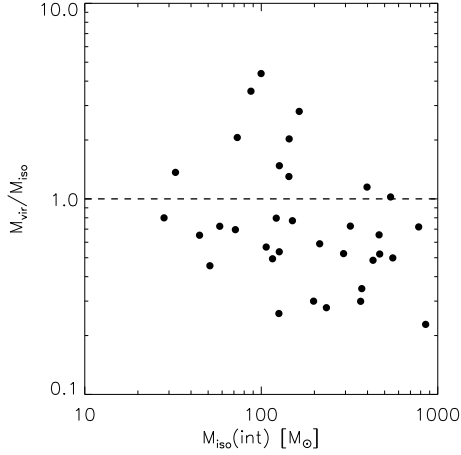} \figcaption{\label{mvirtom1mm}
 Virial parameter (\mvir/\miso) versus \miso(int).  The dashed line marks the mean virial parameter of 1.   
}
\end{figure}

\mvir, \miso(120\as), and \miso(int) 
are given in Columns $2-4$ of Table \ref{derived_masses_densities}. 
For Subset 1, the isothermal masses within a 120\as\ aperture
range from 47 \msun\ to 610 \msun\ with a mean of
280$\pm$180 \msun.  Similarly, \miso(int) ranges from 33 \msun\ to 850 \msun\
with a mean of 300$\pm$220 \msun.  The virial masses range from
33 \msun\ to 560 \msun\, with a mean of
220$\pm$160 \msun.  The distributions of \miso(120\as) and \miso(int) 
are shown in Figure \ref{masshists}.  The mean \mvir\ to \miso(int) ratio is
0.96$\pm$0.93 for Subset 1.  
The ratio of \mvir\ to \miso(int) is shown versus \miso(int) in Figure \ref{mvirtom1mm}.  
The dashed line marks the mean ratio of 1.0.
The trend of decreasing \mvir/\miso\ reflects the dependence of the ratio
on the inverse of the source size at a given distance.  
$R_{obj}$ increases with \miso(int) which lowers \mvir/\miso\ as a 
function of \miso(int).

In these calculations we have made the simplifying assumption that
the density is uniform.  However, if the emission is not spherical or if
the density follows a power-law the virial mass will differ.  A corrected virial 
mass is given by 
\begin{equation} M_{vir}=\frac{5\sigma_v^{2}R}{a_{1} a_{2} G},\label{mvircorrectedeqn}
\end{equation}
\begin{equation} a_{1}=\frac{1-p/3}{1-2p/5}\ \mbox{for}\ p<2.5,
\label{a1}
\end{equation}
(Bertoldi \& McKee 1992) where $a_{1}$  is the correction for a power-law density distribution,
$a_{2}$ is the correction for a non-spherical shape, and $p$ is the density power-law 
index from $n(r)=n_f(r/r_f)^{-p}$ (Mueller et al. 2002).  The mass ratio is then given by
\begin{equation}
\frac{M_{vir}}{M_{iso}}=\frac{\sigma_{v}^2\theta_{rad}B_{\nu}(T_d)\kappa_{\nu}}{2GS_{\nu}Da_1a_2} \propto \frac{\kappa_{\nu}}{Da_1a_2}. \label{massratioeqn}
\end{equation}
The mean aspect ratio of our sources is 
$\langle R_{maj}/R_{min}\rangle = 1.5\pm0.3$, which gives $a_2$ = 1.1.  
To estimate the effect of a power-law density distribution,
we adopt the mean power-law index found by Mueller et al. (2002), 
$\langle p\rangle =1.8\pm0.4$, which was derived by modeling the emission from
31 massive star-forming regions, and find $a_1=1.43\pm0.34$. 
If the corrections for a power-law density distribution (Equation \ref{a1})
and for a non-spherical shape are included, then 
$\langle M_{vir}/M_{iso}\rangle$ would be decreased by 36\% to
0.61$\pm$0.59 for Subset 1.  However, since we do not have accurate 
values of the power-law index for our individual sources, we will 
not include the correction in our calculations or subsequent discussion.   
Magnetic fields are also excluded from our virial mass calculation, 
but if magnetic fields provide additional support to the cloud,
then the true mass would be larger than given by Equations \ref{mvireqn} or 
\ref{mvircorrectedeqn}.

The observed scatter in \mvir\ to \miso\ (Figure \ref{mvirtom1mm})
could be a result of
deviations from the assumed distance of 2.10 kpc or deviations from
Equation \ref{mvireqn} due to different power-law density
distributions or source ellipticities, all of which could differ from
source to source and produce the observed scatter.  Additional
assumptions which could produce the observed scatter include the
assumptions that the 
sources are virialized and that the velocity dispersion measured at
the peak of the mm emission is also valid at $R_{obj}$.  
A virial parameter, \mvir/\miso, of 1 suggests that both mass 
estimates are reasonable. 

\subsection{Volume and Surface Densities\label{avgdensities}}
Average particle density and average surface density can be calculated
based on a mean particle weight per H nucleus ($\mu_p$=2.37)
and a total mass within a given physical size, assuming a uniform
density.  We assumed a spherical volume given by $4\pi R_{obj}^3/3$.  
Mean number densities of particles and mass surface densities were 
calculated for each source (see Column 5 of Table
\ref{derived_masses_densities}).  For Subset 1, \np(int) ranges from 
$1.4\times10^3$ \cmv\ to $17.0\times10^3$
\cmv\ with a mean of $6.2\pm4.4 \times10^3$ \cmv, and the
characteristic values for the full sample are given in
Table \ref{mean_properties}.  
  
The distribution of derived surface densities,
$\Sigma$(int), ranges from 0.019 g \cmc\ to 0.13 g \cmc\ with a mean
of 0.055$\pm$0.032 g \cmc, for Subset 1.
The mean surface density for the full sample is 0.048$\pm$0.030 g \cmc\,
or 230$\pm$140 \msun\ pc$^{-2}$ which is similar to 
the gas surface density threshold for star-formation  of 
200 \msun\ pc$^{-2}$ seen in nearby, low-mass star-forming clouds
(A.~Heiderman et al., in prep.).  This similarity supports the
idea that our BGPS sources are sites of star formation, while
the low volume densities imply we are detecting the less dense, larger
scale portions of clumps rather than the higher density cores.  
  
There is also evidence that a few of the BGPS clumps in this region
include multiple higher density sources when observed with 
interferometers (see 
\S\ref{discussion}).  Eight of the 34 BGPS cores also have water maser
emission (see Table \ref{gas_props}) which requires densities greater
than $10^{10}$ \cmv\ (Strelnitskij et al. 1984; Elitzur et al. 1989),
suggesting that the clumps contain much higher density substructure.

\subsection{Excitation Temperature\label{excitationtemp}}
\begin{figure}[t]
\plotone{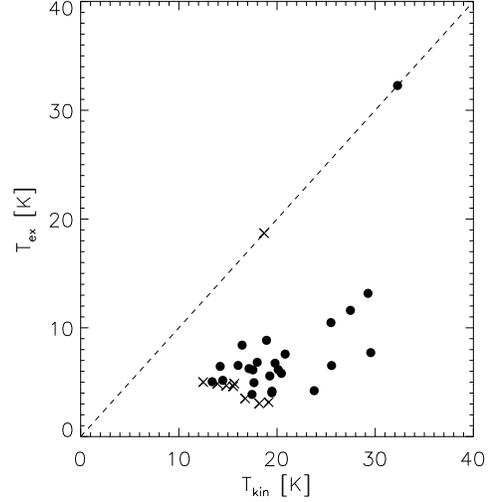} \figcaption{\label{tkin_tex} 
Excitation temperature of the \ammonia(1,1) transition versus kinetic temperature for the full sample.  The filled circles mark subset 1, and crosses mark sources with upper limits in T$_{\mbox{kin}}$.  The dashed line marks a one to one correlation but is not a fit to the data.
}
\end{figure}

The measured excitation temperatures are plotted against the kinetic
temperatures in Figure \ref{tkin_tex}.  The filled circles represent
Subset 1, and the corresponding distribution is characterized by
$\langle$\tex(1,1)$ \rangle =7.8\pm5.6$ K.  There are two sources
where the best fitting model requires the limiting case \tex$=$\tk,
and higher quality data would be needed to constrain \tex.

The mean difference between kinetic and excitation temperatures for Subset 1 is
$\langle$\tk$-$\tex$\rangle = 12.8\pm4.5$ K.  
There are two possible explanations for the differences in derived
temperatures.  The first possibility is inherent in the way we measure
the excitation temperature.  As described in \S\ref{nh3parameterestimation}, 
in order to calculate
\tex\ we must assume something about the ratio of the solid angle of
the source to that of the beam, which here we set to unity.  A lower
excitation temperature could be artificially induced by over-estimating
the beam filling factor.

Another possible explanation is the density in the region.  The level
populations are determined by both collisional and radiative processes
which depend on the local density.  Pavlyuchenkov et al. (2008) study
the conditions under which molecular rotational lines form using
radiative transfer simulations of CO(J=2$-$1) and HCO$^+$(J=1$-$0),
and define a thermalization density $n_{th}$, where \tex\ is within
5\% of \tk.  At densities greater than $n_{th}$, collisions regulate
the level populations by both exciting and de-exciting molecules. 
A commonly used measure is the critical density, defined by 
$n_{crit} = A/\gamma$, where $A$ is the spontaneous emission rate and $\gamma$
is the collision rate coefficient. For low frequency transitions,
such as the \ammonia\ inversion lines, 
collisions are negligible for $n < n_{crit}$, and the
excitation temperature is set by the cosmic background radiation.  At
intermediate densities where $n_{crit} < n < n_{th}$, the excitation
temperature is set by both collisions and radiation and is equal to
neither the background radiation temperature nor the kinetic
temperature.

\begin{figure}[t]
\epsscale{1.2}
\plotone{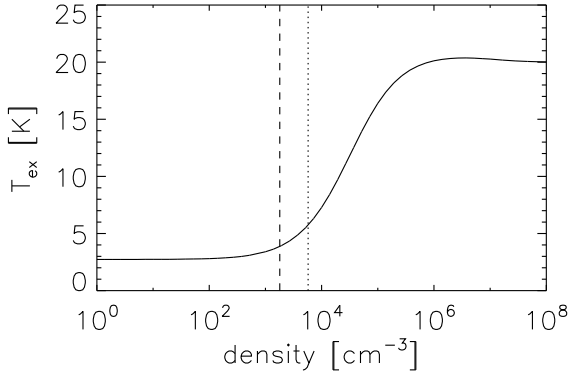} \figcaption{\label{texvsdensity} 
Excitation temperature versus volume density for \ammonia(1,1) for \tk=20 K.  Vertical dashed line denotes $n=n_{crit}$(1,1), and the dotted line denotes the mean particle density found for subset 1.  The LVG code predicts an excitation temperature of 6.3 K at the mean density of our sample.  Our mean excitation temperature is 7.3 K.  
}
\end{figure}

Using an LVG code (Snell et al. 1984), we calculate \tex\ as a 
function of density for a gas temperature of \tk$=20$ K, with  
$N(\ammonia) = 10^{14.5}$ \cmc, over a density range of 100$-10^8$
\cmv.  We assume a line width of 0.75 \kms\ to include radiative trapping.  
Figure \ref{texvsdensity} plots \tex\ as a function of volume density for the
\ammonia(1,1) transition.

From these models we find $n_{th}=210n_{crit}$ for the \ammonia(1,1)
transition.  The large difference in $n_{crit}$ and $n_{th}$ for the
\ammonia\ transitions is due to the low Einstein A coefficient and the
importance of stimulated processes to the level populations.  For
comparison, Evans (1989) finds $n($\tex$=0.9$\tk$)=1.33n_{crit}$ for
the CS(3-2) transition.  Although this criterion is less stringent
than $n_{th}$, it demonstrates the differences between the low
frequency radio transitions and higher frequency submillimeter
transitions.  The low frequency transitions, such as the
\ammonia\ transitions, require a density much higher than the critical
density before \tex\ begins to approach \tk.

Although the volume-averaged densities derived from the millimeter
continuum are approximately equal to the critical densities for both
the \ammonia(1,1) and (2,2) transitions ($1.8\times10^{3}
\mbox{cm}^{-3}\ \mbox{and\ } 2.1\times10^{3}\ \mbox{cm}^{-3}$
respectively; Tielens 2005), the densities are not high enough for the
gas to be completely thermalized, and \tex$<$\tk\ as expected at
densities near the critical density.

\subsection{\ammonia\ Excitation Density\label{np_nh3}}
If we assume the gas is not in LTE, we can define the excitation
density, $n_{ex}$, as the density required to produce level
populations described by \tex\ for a given gas kinetic temperature.
We calculate $n_{ex}$ from \tk, \tex, and $\tau$ via the balance of a
two-level system.  The excitation density is given by (Foster et
al. 2009; Swift et al. 2005; Caselli et al. 2002)
\begin{equation} n_{ex} = \frac{k(J(\tex)-J(T_{cmb}))}{h \nu_{(1,1)}(1-J(\tex)/J(\tk))}n_{crit}\beta, 
\label{nexeqn}
\end{equation}
where
\begin{equation}
J(T)=\frac{h \nu_{(1,1)}}{k(1-e^{-h\nu_{(1,1)}/kT})}.
\label{jeqn}
\end{equation}
$T_{cmb} = 2.73$ and the escape probability is given by $\beta = (1-
e^{-\tau})/\tau$, where we take $\tau = \tau_{(1,1)}\times 0.233$
which is the maximum optical depth in a single hyperfine line of the
\ammonia(1,1) transition.

\begin{figure}
\epsscale{1.1}
\plotone{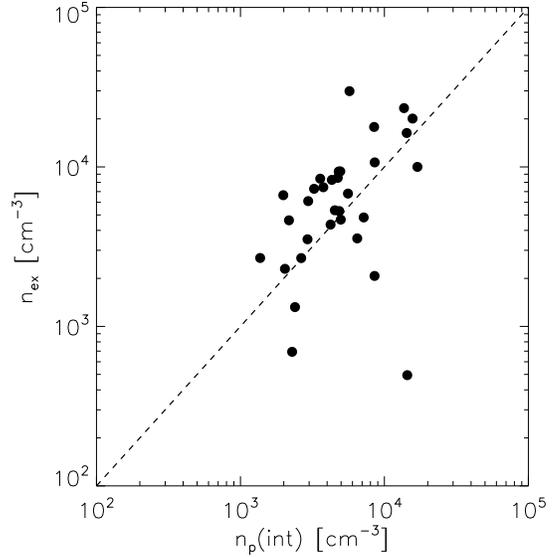}\figcaption{\label{npnh3fig}
\ammonia\ excitation density, $n_{ex}$, versus \np(int).  The dashed line marks a one-to-one correlation and is not a fit to the data.  The mean volume densities are in good agreement with the \ammonia\ excitation densities.
}
\end{figure}

The excitation densities are listed in Column 6 of 
Table \ref{derived_masses_densities}.
We cannot calculate $n_{ex}$ for the two sources
where we set \tex=\tk\ (see \S\ref{excitationtemp}).  The excitation densities
for Subset 1 range from $2.1\times10^3$ \cmv\ to $29.9\times10^3$
\cmv\ with a mean of $9.6\pm7.1\times10^3$ \cmv.  
Figure \ref{npnh3fig} shows a comparison of the \ammonia\ excitation
density and the volume-averaged density (\np) from the millimeter data.  
The solid line marks a one-to-one correlation, and
is not a fit to the data.  Although there is considerable spread, 
\np\ is in good agreement with the excitation density.
The mean ratio of \np(int) to excitation density is
$1.7\pm1.0$ with a median of $1.8$ for subset 1.  The agreement between
$n_{ex}$ and \np\ provide additional support that the \ammonia\ 
and continuum emission trace the same physical region.  

\subsection{Column Densities\label{coldensection}}

The column density per beam of \ammonia\ in the (1,1) state
can be calculated from the observed parameters
via the following equation:
\begin{equation} N(1,1)=2\ \frac{8 \pi \nu_{(1,1)}^2}{c^2} \frac{g_1}{g_2} \frac{1}{A_{(1,1)}} \left[ 1-e^ {  \frac{h \nu_{(1,1)}}{k T_{ex}}} \right]^{-1}\int\tau(\nu)d\nu,
\label{N11}
\end{equation}
where
\begin{align} \int \tau(\nu)d\nu =& \tau(1,1) \int d\nu\ \sum_i s_i\ \exp^{-\frac{(\nu-\nu_i-\nu_{v_{LSR}})^2}{2\sigma_i^2}} \notag \\
=& \sqrt{2\pi}\frac{\sigma_v}{c}\nu_{(1.1)} \tau(1,1)
\label{taunu}
\end{align}
 and $\nu_{(1,1)}$ is the central frequency of the (1,1) transition,
$\tau_{(1,1)}$ is the total optical depth in the (1,1) transition, $g_1$ and $g_2$ are the statistical
weights of the upper and lower levels of the (1,1) inversion transition and are equal, 
$A_{(1,1)} = 1.68\times10^{-7}\ \mbox{s}^{-1}$ is the 
Einstein A coefficient for the transition (Pickett et al. 1998), and $\sigma_v$ is the Gaussian $1/e$ line width.
The leading factor of 2 in equation \ref{N11} assumes an equal number of \ammonia\
molecules in each state of the inversion transition, which is justified since
the states have equal statistical weight and are separated in energy by 0.05 K.
The total optical depth in the (1,1) transition is split over the 18 hyperfine components, which
is expressed in the summation in Equation \ref{taunu}.  $s_i$ is the weight of each hyperfine 
component, $\nu_i$ is the frequency of each hyperfine component, and $\sigma_i$ is the 
Gaussian $1/e$ width of each hyperfine component.

The total \ammonia\ column density, $N_{NH_3}$, can be calculated by
scaling $N(1,1)$ by $Z/Z(J=1)$ where Z is the partition function given
in Equations \ref{partitionfunctionO} and \ref{partitionfunctionP},
$Z(J=1)$ is the population in only the $J=1$ level, and we truncate
the summation at 50 terms.  The \ammonia\ column densities are given
in column 8 of Table \ref{derived_masses_densities}.  
$N_{NH_3}$ ranges from $4.2\times10^{13}$
\cmc\ to $75.8\times10^{13}$ \cmc\ with a mean of $3.5\times10^{14}$
\cmc\ for subset 1.

We can also calculate the column density of H$_2$ via the millimeter
dust observations.  From the gas and dust mass within a given
aperture, we can calculate $N_{H_2}$ by
\begin{equation}
N_{H_2} = \frac{\miso}{\mu_{H_2} m_H \pi R_{obj}^2},
\label{h2colden}
\end{equation}
where $\mu_{H_2}=2.8$.  $N_{H_2}$ is listed in column 9 of Table 
\ref{derived_masses_densities}.  $N_{H_2}$
ranges from $4.1\times10^{21}$ \cmc\ to $27.1\times10^{21}$ \cmc, with
a mean of $1.2\pm0.7\times10^{22}$ \cmc.  Using the relationship
between H$_2$ column density and visual extinction of Bohlin et
al. (1978) assuming $R_V = 3.1$, $N_{H_2}/A_V=9.4\times10^{20}$
\cmc\ mag$^{-1}$, we find
$A_V=12.4$ mag for the mean H$_2$ column density of our sample.
However, $R_V = 3.1$ may not be an accurate description of the
dust in these dense regions.  Chapman et al.~(2009) find
that regions with $A_K > 1$ mag ($A_V > 5$ mag) are described
well by a $R_V = 5.5$ extinction model (Weingartner \& Draine 2001).
Using this extinction model, $N_{H_2}/A_V = (2 \times 1.086 C_{ext})^{-1}$,
where the factor of two converts $N_H$ to $N_{H_2}$ and $C_{ext}$ is
the extinction cross section in the V band.  Using 
$N_{H_2}/A_V = 6.86\times10^{20}$ \cmc\ mag$^{-1}$ we find 
$A_V=17.5$ for the mean column density of our sample.

We can obtain a measurement of the \ammonia\ abundance by comparing
the column densities of \ammonia\ and H$_2$.  The abundances are
presented in column 10 of Table \ref{derived_masses_densities}.  
The abundance ranges
from $0.89\times10^{-8}$ to $8.3\times10^{-8}$ with a mean of
$3.1\times10^{-8}$ for subset 1.  These
results are similar to the values derived for dense cores,
e.g. Harju et al.~(1993, $3\times10^{-8}$), 
Tafalla et al.~(2006, $2.8\times 10^{-8}$) 
and Foster et al. (2009, $2\times 10^{-8}$).

\section{Discussion}\label{discussion}

The physical properties of the BGPS sources in the Gemini OB1 Molecular
Cloud show that the majority of the sources are the clumps from which
entire stellar clusters will form.  
Characteristics of all properties are listed in Table \ref{mean_properties} 
for both the full sample and subset 1, where numbers in parentheses are 
the standard deviation rather than an error in the mean.   Recall that clumps are typically
characterized by masses ranging from $50-500$ \msun, radii of $0.3-3$ pc,
mean densities of $10^3-10^4$ \cmv, and gas temperatures of $10-20$ K 
(Bergin \& Tafalla 2007).  In the Gemini OB1 region, we find a mean
isothermal mass of $\langle$\miso(int)$\rangle=200\pm220$ \msun, a mean radius
of $\langle$R$\rangle=0.56\pm0.19$ pc, a mean density of 
$\langle$\np(int)$\rangle=6.2\pm4.4 \times10^3$ \cmv, and a mean
gas kinetic tempearture of $\langle$\tk$\rangle=20.6\pm5.1$ K.  The mean
properties of our BGPS sources fall squarely in the ranges of 
characteristic clump properties.

In contrast, the unresolved BPGS source in this region (ID \#51) has 
properties 
within the range of the characteristic properties of cores.  Recall
that cores have masses from $0.5-5$ \msun, sizes of $0.03-0.2$, mean 
densities of $10^4 - 10^5$, and gas temperatures between 8 and 12 K 
(Bergin \& Tafalla 2007).  Source \#51 has \miso(int)$=44\pm13$ \msun,
an upper-limit for size of R$< 0.17$ pc, a mean density of 
\np(int)$=2.9\times10^3$ \cmv, and an upper-limit of gas kinetic
temperature of \tk$<14.8$ K.  \miso(int) is significantly above
the range of core masses, but also falls below the 
range of characteristic clump masses.  The mean density does not fall 
within the characteristic density range of cores, but the upper-limit
used for the source size could be the explanation.  An unresolved source must
have a size smaller than the beam size, so the beam size is
artifically giving a lower density.  There are additional
unresolved BPGS sources but they do not have corresponding \ammonia\
pointings and were excluded from our analysis.

Further evidence that the BGPS sources are not directly forming 
a single massive star is in the mean surface density of the sample, 
$\langle\Sigma(int)\rangle=0.055\pm0.032$ g cm$^{-2}$.  Krumholz
\& McKee (2008) have shown that a surface density of at least
1 g cm$^{-2}$ is required to avoid fragmentation and form a single massive
star.  Although the mean masses of our clumps are high enough to form
a massive star given typical star formation efficiencies, the
mean surface density is not high enough to prevent fragmentation and
suggests that there should be significant substructure within these
clumps.  Substructure has been seen in higher resolution studies of
some of the Gemini OB1 BGPS clumps and is discussed in detail in the 
following section.  
Additionally, the mean virial parameter for Subset 1 is approximately
1 which suggests that the BGPS sources are stable against collapse
on the size scales the survey is sensitive to.

\subsection{Comparison to Other Studies}\label{comparison}

A comparison to the literature provides further evidence that the BGPS sources are
likely the clumps from which stellar clusters will form.  
The BGPS sources often resolve into multiple smaller, higher density regions when 
observed at higher resolution with telescopes such as the SMA, SCUBA on the JCMT,
and SHARC on the CSO.  Densities calculated from 1.3 mm continuum SMA observations
are on the order of $10^6 - 10^7$ \cmv\ (Cyganowski et al. 2007), 
a factor of $10^3$ greater than the densities of the BGPS sources presented here.
Models of observed \sed s require both a hot ($\sim100$K) and warm ($\sim40$K)
dust component (Minier et al. 2005), again suggesting that the 
BGPS sources are the cooler, lower density clumps which contain hotter cores.  
When comparing our 1.1 mm emission to the large-scale CO emission, we find that 
the BGPS sources have H$_2$ column densities a factor of 10 greater than the mean
column density of the regions of the molecular cloud without star formation
(Carpenter et al. 1995a).  
The resolved BGPS sources are typically clumps which are
embedded in a molecular cloud and contain higher density, warmer cores while the
unresolved BGPS sources are likely to be cores.   Detailed comparisons for a few sources 
and the entire molecular cloud are presented below.

\subsubsection{S252}
S252 corresponds to G188.948$+$00.883 (ID \#26) in the BGPS.  Minier et al. (2005; hereafter
M05) observed methanol and C$^{\mbox{18}}$O toward S252, and also modeled mid-IR
to millimeter wavelength emission.  Using SCUBA on the James Clerk Maxwell Telescope and 
SIMBA on the Sweedish ESO Submillimetre Telescope (SEST), they resolve S252 into
two millimeter sources, G188.95$-$MM1 and $-$MM2.  Models fit to the spectral energy distributions
require a warm extended envelope (42 and 50 K) for each source and a hot central
component (150 K) for MM1.  The rotational temperature derived from their methanol
observations is 58 K.  The temperatures derived from both SED fitting and molecular line
observations are all significantly warmer than our \ammonia\ kinetic temperature of 29 K.
In addition to warmer temperatures, M05 also derived a smaller mass than the BGPS.
The BGPS does not resolve two sources here, and finds \miso(int) =556 \msun, while 
M05 finds a total mass for MM1 and MM2 of 155 \msun\ when scaled to our assumed
distance of 2.1 kpc.

\subsubsection{S252A}
Mueller et al. (2002; hereafter M02) observed S252A in 350\um\ continuum emission using
SHARC at the CSO, and modeled the observed SED including wavelengths ranging
from 12 \um\ to 1.1 mm.  They fit a power-law density distribution and determined a density
normalized at a radius of 1000 AU, $n_f$.  S252A in the M02 sample corresponds to
G189.776$+$00.343 (ID \#13) in our sample.  M02 calculated an isothermal dust
mass of 220 \msun\ within a 120\as\ aperture when scaled to our assumed distance.
We find \miso(120\as)$=373\pm40$\msun from the 1.1 mm emission.  
SHARC filtered out extended emission on a smaller scale than the BGPS,
so the M02 observations detected the warmer, higher density, and lower mass 
substructure within the BGPS source.  
The $T_{iso}$ given by M02 was 30K, marginally warmer than the temperature derived 
from the \ammonia\ observations, \tk$=26$K.  

In general, the M02 study included higher-mass, warmer sources than our sample.  
Their sample selection required the presence of strong water maser emission 
(Arcerti water maser catalog; Cesaroni et al.~1988) as a signpost of massive
star formation.
M02 found $\langle$\miso(120\as)$\rangle=2040\pm4410$
\msun, with a median of 400 \msun\ for their sample of 51 massive star-forming regions.  
In contrast, we find
$\langle$\miso(120\as)$\rangle=280\pm180$ \msun, with a median of 190
\msun.  Our sources are approximately a factor of 10 less massive than
the sample studied by M02.

\subsubsection{S255N and S255IR}
S255N corresponds to G192.581-00.043 (ID\#45) in our sample, while S255IR
corresponds to G192.596-00.051 (ID \#47).  We find \miso(int)$=397\pm32$ 
\msun\ and \tk$=30$K for S255N, and \miso(int)$=540\pm43$ and \tk$=32$K
for S255IR.  We compare our BGPS and \ammonia\
derived properties with studies by Zinchenko et al. (1997; hereafter Z97), M05, and 
Cyganowski et al. (2007; hereafter C07) in detail.

Z97 mapped 
\ammonia(1,1) and (2,2) emission from S255 with the 100~m Effelsberg radio telescope.  
They centered their map
on an IRAS point source located between S255N and S255IR, and considered the emission
from both to be part of a single source.  They report observed values at two locations,
one of which corresponds very well with S255N and our source G192.583-0.043.  The observed
\ammonia\ properties between the two studies agree fairly well considering the \ammonia\ 
positions differ slightly.  For example, Z97 find 
$v_{lsr}=8.65$ \kms\ and $\sigma_v=\Delta v/(8 \ln 2)^{1/2}=1.02$ \kms, while we find 
$v_{lsr}=8.66$ \kms\ and 
$\sigma_v=1.31$ \kms.  Z97 find that in half of their sources
the \ammonia\ line width increases toward the \ammonia\ peak
which could account for the same radial velocity but different line widths.  
Both studies derive slightly different \tk, with Z97 finding \tk=23 K and our study finding
\tk=30 K for S255N and 32 K for S255IR.  The Z97 mass derived from the
 \ammonia\ column density and a canonical abundance
is 420 \msun\ when scaled to our assumed distance.  Since this includes both 
S255N and S255IR, this mass should be compared
to the sum of masses for the two BGPS sources, \miso(int,tot) $=937$ \msun.
Our total \miso(int) is comparable to the virial mass Z97 derive, \mvir(Z97)
$=1134$ \msun\ when scaled to the maser parallax distance.

C07 observed S255N in 350
\um\ continuum with SHARC on the CSO and 1.3 millimeter continuum with
the SMA.  S255N-MM1 corresponds to G192.581$-$0.043 in our sample (ID
\#45).  SHARC ($\Theta_{mb}=$15\as) recovers only a single peak at the
position of S255N-MM1, while the SMA (4.7\as$\times$2.4\as\ beam)
clearly resolves three separate sources: SMA1, SMA2, and SMA3.  The
total beam-averaged mass of all three sources scaled to the parallax
distance is 17$-$42 \msun, where the range is due to uncertainty in
the assumed dust temperature.  The assumed temperature range for SMA1
is 40$-$100 K, while the other two sources are likely between 20 and
40 K based on the presence of molecular emission in SMA1 and the lack
thereof in SMA2 and SMA3.  We find \tk$ = 29.5$ K for source \#45 in
our sample, which is consistent with the range of temperatures for
SMA2 and 3 and is just below the lower end of the temperature range
for SMA1.  The difference in the temperatures reflects the different
material probed with the two sets of observations: the SMA observes
the warm, high density structure within the cooler, lower density
clumps to which the BGPS and GBT observations are sensitive.

C07 derived densities in the range of $10^6-10^7$ \cmv\ from the SMA
1.3 mm continuum emission.  These densities are $\sim10^3$ times higher than
the derived BGPS densities.  With a smaller beam size, the SMA
resolves out the low density material which the BPGS is sensitive to
and is able to detect the smaller, higher density regions within the
BGPS clump.  Our calculated \np\ is a volume averaged density which
will underestimate the peak density of the regions where massive stars
will form because a larger percentage of the volume is at the lower
densities.

Similar to C07, M05 found that S255N requires a hot component of 106 K
and a warm component of 44 K while S255IR requires a hot component of
225 K and a warm component of 42 K based on fitting models to the observed 
\sed.  M05 find \miso\ of 140 and 210 \msun\ for S255N and S255IR, respectively.
Their masses are significantly smaller than our isothermal masses, which
is consistent with the paradigm in which observations with a smaller beam size
will resolve out the lower density emission and detect the substructure 
within the BGPS sources.  

Chavarr\`ia et al.~(2008) provides direct evidence that S255IR and 
S255N are forming a cluster.  They used \textit{Spitzer} IRAC and NIR data
to identify sources with near- or mid-IR excesses.  They identified
a total of 10 clusters of YSOs as well as a population of distributed
YSOs.  The largest cluster they identified corresponds to both
S255IR and S255N (ID\#45 and \#47 in our sample).  They detected
140 stars with an IR excess, and assume the total cluster membership to
be 488 sources after accounting for completeness and sources with
no IR excess.  

\subsubsection{S258 and G192.719$+$00.043}
There is also direct evidence of clusters forming in both
S258 (G192.644$+$00.003; ID \#48) and G192.719$+$00.043 (ID \#49)
Chavarr\`ia et al.~(2008) identified  49 sources with IR excesses
in S258, and estimate the total number of cluster members to be 
186 after considering completeness limits and the number of 
potential sources without an IR excess.  The cluster they 
identified in G192.719$+$00.043 includes 22 detected members
and is estimated to include 44 total members.

\subsubsection{Gemini OB1 Molecular Cloud}\label{carpenter95a}
Carpenter et al. (1995a; hereafter C95a) surveyed 32 square degrees of
the Gemini OB1 molecular cloud complex in $^{12}$CO($J=1-0$) and
$^{13}$CO($J=1-0$).  The detailed CO maps yield a wealth of physical
properties, which provide an independent check of the properties
determined in this study.  C95a separated the molecular cloud into
seven subregions marked by bright extended $^{13}$CO emission.
Subregions 1$-$3 are associated with optical HII regions and
correspond to groups 1-3 in this paper.  Subregion 6 contains a
Herbig Ae/Be star HD250550, but is not included in our BGPS image.
The remaining subregions have no optical evidence of star formation.

Along the lines of sight with a 5$\sigma$ detection in $^{13}$CO, C95a assume
the $^{12}$CO is optically thick and can determine \tk.  For subregions 4$-$7 with
no star formation, \tk\ is typically $6-8$ K.  In contrast, in subregions
$1-3$ with active star formation and optical HII regions the bulk of the material is found
to be $6-8$ K but the distribution has a warm tail reaching up to \tk$= 25$ K.
The kinetic temperatures ($\sim10-30$ K) we derive from the \ammonia\ spectra
are consistent with the warmer CO lines of sight where active star formation is 
likely occurring.  
C95a also determines the column density of H$_2$ along lines of sight with a 5$\sigma$
detection of $^{13}$CO.  The subregions without star formation have
$\langle N_{H_2} \rangle = 1.0-1.3\times 10^{21}$ \cmc, while the cloud as a whole
ranges from $1.2\times10^{20}$ \cmc\ to $6.4\times10^{22}$ \cmc.  This is again consistent
with our study, where we find $\langle N_{H_2}(int) \rangle = 1.2\times10^{22}$ \cmc.

For the subregions which correspond to our groups, we can determine the fraction of mass
in dense clumps.  To do so, we compare \miso(int) with $M_{LTE}$ from C95a.  The total
mass  of dense millimeter clumps is 1835, 4549, and 1973 \msun\ for groups $1-3$, respectively.
The corresponding $M_{LTE}$ for subregions $1-3$ are $3.5\times10^4, 5.2\times10^4$, and
$4.3\times10^4$ \msun.  We find 5.2\%, 8.7\%, and 4.6\% of the total mass is in the dense
clumps for subregions and groups $1-3$, respectively.  
C95a find that 10\% of the 
molecular mass traced by CO is contained in lines of sight with H$_2$ 
column densities greater than $10^{22}$ \cmc, similar to the mean column
density of our clumps.  This percentage is also similar to nearby,
low-mass, star-forming molecular clouds Perseus, Ophiuchus, and Serpens, where 
$<10\%$ of the total mass is in dense cores (Enoch et al. 2007).
The fraction of mass found in the dense regions associated with star formation
is very similar using different tracers as well as in high-mass and low-mass 
star-forming regions, two very different environments.

\subsection{What is a BGPS source?\label{bgpssource}}
The BGPS is detecting thousands of sources which span a vast range of
distances, and it is important to understand the biases and trends
inherent in the data.  In contrast to the inner Galaxy BGPS sources,
the Gemini OB1 molecular cloud provides an excellent opportunity to
characterize the clump properties without large uncertainties in
distances.  Two neighboring sources in the inner Galaxy could be many
kpc apart along the line of sight, whereas the Gemini OB1 sources are
all located at the same distance.  Since we have an accurate measure
of the distance to the Gemini OB1 sample we will use it as a starting
point to understand the effects of distance on the derived properties.

The detection criterion for a BGPS source is equivalent to a
5$\sigma$ detection within an area the size of a beam (Rosolowsky et
al., 2009).  This results in a minimum mass per beam, below which the
emission will be lost in the noise or removed during the sky cleaning
process.  This minimum mass is given by Equation \ref{miso}, with
\Snu=$5\sigma$, where $\sigma$ is the RMS noise per pixel in units of
Jy~beam$^{-1}$ in the BGPS image, $D$ is the distance, $\kappa_{\nu}$
is the dust opacity per gram of gas and dust, and $B_{\nu}(T_d)$ is
the Planck function evaluated at the dust temperature.  Thus the BGPS
is subject to the Malmquist bias: as the distance increases, the
minimum mass to which the BGPS is sensitive also increases.

The minimum volume-averaged particle density to which the BGPS is
sensitive is given by
\begin{equation} n_{p,min}=\frac{15\sigma}{4\pi\mu m_H \kappa_{\nu} B_{\nu}(T_d) \Theta_{mb}^3 D}\propto \frac{1}{\Theta_{mb}^3 D},
\label{nmin}
\end{equation}
where $\mu=2.37$, $m_H$ is the mass of a hydrogen atom, and
$\Theta_{mb}$ is the FWHM of the CSO beam at 1.1 mm (33\as).  As
 the distance increases, the minimum mass increases (Equation
\ref{miso}), and the minimum density decreases (Equation \ref{nmin})
for sources smaller than the beam size, quantifying how BGPS
observations are subject to Malmquist bias and beam dilution respectively.
Similarly, if we allow the aperture size for photometry to vary at a constant
distance, larger sources will have lower densities.

For the Gemini OB1 molecular cloud observations, $\sigma=0.37$
Jy/beam, requiring \Snu(beam)$ = 0.35$ Jy for a 5$\sigma$
detection.  The corresponding minimum mass is $M_{min}=20$ \msun\ at
a distance of 2.1 kpc and \td=20 K.  Thus the minimum density required
to detect a beam-sized source after deconvolution 
in the Gemini OB1 molecular cloud is
$n_{p,min}\approx10^4$ \cmv.  Sources larger than the beam size can have
densities lower than $10^4$ \cmv.  The smallest BGPS sources
could have the highest densities and may be the cores which will form
individual stars, multiple star systems, or small groups, whereas the
larger sources will have generally lower densities and are likely to
be the clumps which will form stellar clusters.

In the inner Galaxy, the largest distance will greatly exceed that of
the Gemini OB1 molecular cloud and the minimum density will be
significantly lower.  At distances greater than approximately 2 kpc,
the BGPS is only able to detect clumps which will form stellar
clusters, and not individual cores.  Since the Gemini OB1 molecular
cloud lies at 2.1 kpc, we may detect some cores as well as
clumps.

\section{Summary}\label{summary}
We present the image and catalog of the Gemini OB1 Molecular cloud
complex from the 1.1 mm Bolocam Galactic Plane Survey.  We observed
\ammonia(1,1), (2,2), and (3,3) inversion transitions and 22 GHz H$_2$O
maser transitions toward 37 BGPS sources, as well as \ammonia(4,4) 
and C$_2$S toward 10 sources with the GBT. We have obtained a detection $\geq 5\sigma$
in the (1,1) transition of 34 of the 37 BGPS sources, and a
detection $\geq 5\sigma$ in the (1,1) and (2,2) transitions in 25 of the 34 BGPS
sources.  

We use a seeded watershed algorithm to extract the millimeter continuum 
sources from the BGPS maps.  We find a mean FWHM of sources 
of 110\as, or 1.12 pc at a distance of 2.1 kpc.
We estimate the gas physical properties by fitting a simple model to
all \ammonia\ transitions simultaneously.
From these fits, we find $\langle$\tk$\rangle=20$ K, and velocity dispersions ranging
from $0.4-1.3$ \kms.  We calculate the non-thermal velocity dispersion,
compare it to the expected thermal sound speed, and find a ratio of
$\sigma_{NT}$ to $a$ ranging from 1.5 to 4.2 with a mean of
$2.6\pm0.8$.  The velocity dispersions of all sources are dominated by
non-thermal motions.

We calculate the isothermal mass for each millimeter source
based on \Snu(120\as) and \Snu(int) assuming
$\td=\tk$ from the \ammonia\ observations.  We find
$\langle$\miso(120\as)$\rangle=230\pm180$ \msun\ and
$\langle$\miso(int)$\rangle=250\pm210$ \msun, where the error is the
standard deviation of the sample.  We also calculate the virial mass
within the source size derived by the source extration algorithm under the assumption
that the \ammonia\ emission traces the millimeter emission well.  We find
$\langle$\mvir$\rangle=190\pm160$ \msun.  The mean \mvir\ to
\miso\ ratio is 1.0$\pm$0.95.  The total isothermal mass for the portion of
the Gemini OB1 molecular cloud we have mapped is 8,400 \msun, which 
represents 6.5\% of the total mass seen in CO observations.

We also calculate the volume-averaged densities and find
$\langle$\np(int)$\rangle=5.9\pm4.3\times10^3$ \cmv, with the
highest density being \np$_{,max}$(int)$=1.7\times10^4$ \cmv.  The
mean surface density of the full sample is
$\langle\Sigma$(int)$\rangle=0.048\pm0.030$ g cm$^{-2}$.
We calculate both \ammonia\ and H$_2$ column densities and find mean
values of $3.0\times10^{14}$ \cmc\ and $1.0\times10^{22}$ \cmc,
respectively.  We find a mean \ammonia\ abundance of
$3.0\times10^{-8}$ relative to H$_2$.

The mean sizes, densities, kinetic temperatures and surface densities
of the BGPS sources
suggest they are cluster-forming clumps that contain the
cores that will form individual or small multiple systems.

\acknowledgements
We thank the anonymous referee, whose comments and suggestions have
greatly improved this paper.
The Caltech Submillimeter Observatory (CSO) is operated by Caltech
under a contract from the National Science Foundation.  The Green Bank
Telescope (GBT) is operated by the National Radio Astronomy
Observatory. The National Radio Astronomy Observatory is a facility of
the National Science Foundation, operated under cooperative agreement
by Associated Universities, Inc.  We would like to acknowledge the
staff and day crew of the CSO, and also the GBT operators and staff
for their assistance.  The BGPS is supported by the National Science
Foundation through NSF grant AST-0708403.  The first observing runs
for the BGPS and the GBT observing runs were supported by travel funds
provided by the National Radio Astronomy Observatory (NRAO).  J.A. was
supported by a Jansky Fellowship from the NRAO.  Support for the
development of Bolocam was provided by NSF grants AST-9980846 and
AST-0206158.  M. K. Dunham and N. J. Evans were supported by NSF grant
AST-0607793 to the University of Texas at Austin.  E.R. was supported
by NSF grant AST-0502605 and a Discovery Grant from NSERC of Canada.
C.J.C. was partially supported during this work by a National Science
Foundation Graduate Research Fellowship.  This research has made use
of NASA Astrophysics Data System (ADS) Abstract Service and of the
SIMBAD database, operated at CDS, Strasbourg, France.


\clearpage

\clearpage


\LongTables	
\clearpage

\clearpage
\begin{landscape}
\begin{deluxetable}{rcccccccccccc}
\tabletypesize{\scriptsize}
\tablewidth{0pt}
\tablecaption{\label{nh3_props}Observed NH$_{3}$ Properties}
\tablehead{
\colhead{ID} & \colhead{RA} & \colhead{Dec} & \colhead{V$_{LSR}$} & \colhead{$\sigma$(1,1)} & \colhead{T$_{\mbox{MB}}$(1,1)} & \colhead{W(1,1)} & \colhead{T$_{\mbox{MB}}$(2,2)\tablenotemark{b}} & \colhead{W(2,2)\tablenotemark{b,c}} & \colhead{T$_{\mbox{MB}}$(3,3)\tablenotemark{b,c}} & \colhead{W(3,3)\tablenotemark{b,c}} & \colhead{T$_{\mbox{MB}}$(4,4)\tablenotemark{b,c}} & \colhead{W(4,4)\tablenotemark{b,c}} \\                    
\colhead{Number\tablenotemark{a}} & \colhead{(J2000)} & \colhead{(J2000)} & \colhead{(km s$^{-1}$)} & \colhead{(km s$^{-1}$)} & \colhead{(K)} & \colhead{(K km s$^{-1}$)} & \colhead{(K)} & \colhead{(K km s$^{-1}$)} & \colhead{(K)} & \colhead{(K km s$^{-1}$)} & \colhead{(K)} & \colhead{(K km s$^{-1}$)} 
}
\startdata
01........ & 06 07 30.8 &  20 39 46.1 &  8.60 (0.05) & 0.72 (0.06) & 0.76 (0.16) &  2.39 (0.22) & $<$ 0.79 & $<$ 0.44 & $<$ 0.86 & $<$ 0.48 & \nodata & \nodata \\ 
03........ & 06 07 47.7 &  20 39 31.5 &  7.43 (0.02) & 0.58 (0.02) & 1.69 (0.15) &  5.06 (0.21) & 0.59 (0.15) & 0.59 (0.08) & $<$ 0.81 & $<$ 0.45& \nodata & \nodata \\
04........ & 06 08 18.2 &  20 36 36.1 & 10.17 (0.02) & 0.49 (0.03) & 1.20 (0.16) &  3.18 (0.22) & $<$ 0.80 & $<$ 0.45 & $<$ 0.83 & $<$ 0.46 & \nodata & \nodata \\ 
05........ & 06 08 19.8 &  21 22 10.4 &  2.48 (0.02) & 0.49 (0.02) & 1.66 (0.16) &  4.89 (0.22) & $<$ 0.83 & $<$ 0.46 & $<$ 0.85 & $<$ 0.47 & \nodata & \nodata \\ 
06........ & 06 08 23.5 &  20 36 32.9 & 10.12 (0.02) & 0.51 (0.02) & 0.80 (0.08) &  2.39 (0.11) & 0.29 (0.08) & 0.31 (0.04) & $<$ 0.84 & $<$ 0.46& $<$ 0.44 & $<$ 0.24 \\
07........ & 06 08 25.3 &  20 42 03.3 &  8.69 (0.01) & 0.48 (0.01) & 1.75 (0.08) &  5.15 (0.12) & 0.43 (0.08) & 0.47 (0.05) & $<$ 0.87 & $<$ 0.48& $<$ 0.44 & $<$ 0.24 \\
11........ & 06 08 33.3 &  20 34 38.7 &  8.18 (0.05) & 1.13 (0.05) & 1.01 (0.16) &  4.43 (0.22) & 0.69 (0.16) & 0.70 (0.09) & $<$ 0.80 & $<$ 0.44& \nodata & \nodata \\
12........ & 06 08 31.0 &  20 26 44.5 &  5.49 (0.01) & 0.42 (0.01) & 2.56 (0.16) &  8.38 (0.22) & 0.69 (0.15) & 0.72 (0.08) & $<$ 0.78 & $<$ 0.43& \nodata & \nodata \\
13........ & 06 08 35.2 &  20 39 05.6 &  9.11 (0.01) & 0.87 (0.01) & 2.91 (0.16) & 16.27 (0.22) & 1.72 (0.15) & 4.06 (0.08) & 1.17 (0.16) &  2.09 (0.09) & \nodata & \nodata \\
B & 06 08 29.3 &  20 40 28.5 &  7.82 (0.02) & 0.54 (0.03) & 1.17 (0.15) &  3.78 (0.21) & 0.66 (0.15) & 0.83 (0.08) & $<$ 0.79 & $<$ 0.44 & \nodata & \nodata \\
16........ & 06 08 40.6 &  20 37 56.9 &  7.48 (0.01) & 0.78 (0.01) & 4.15 (0.16) & 18.21 (0.22) & 1.45 (0.15) & 2.67 (0.08) & 0.63 (0.16) &  0.93 (0.09) & \nodata & \nodata \\
17........ & 06 08 40.4 &  21 31 03.1 &  2.60 (0.01) & 0.92 (0.02) & 2.45 (0.16) & 11.35 (0.22) & 1.55 (0.16) & 2.82 (0.09) & 0.72 (0.16) &  1.33 (0.09) & \nodata & \nodata \\
18........ & 06 08 42.1 &  20 36 19.7 &  7.69 (0.01) & 0.71 (0.01) & 2.71 (0.15) & 11.60 (0.21) & 0.96 (0.15) & 1.52 (0.08) & $<$ 0.77 & $<$ 0.43& \nodata & \nodata \\
19........ & 06 08 43.4 &  20 33 59.7 &  8.44 (0.02) & 0.91 (0.03) & 0.81 (0.08) &  3.52 (0.11) & 0.42 (0.08) & 0.57 (0.04) & $<$ 0.84 & $<$ 0.46& $<$ 0.43 & $<$ 0.24 \\
20........ & 06 08 43.9 &  20 32 29.3 &  8.36 (0.02) & 0.76 (0.02) & 1.01 (0.08) &  3.81 (0.12) & 0.38 (0.08) & 0.57 (0.05) & $<$ 0.82 & $<$ 0.45&0.37 (0.09) & 0.25 (0.05) \\
21........ & 06 08 44.4 &  20 37 53.1 &  8.61 (0.01) & 1.01 (0.01) & 2.43 (0.09) & 12.30 (0.12) & 0.83 (0.08) & 1.56 (0.05) & $<$ 0.87 & $<$ 0.48& $<$ 0.43 & $<$ 0.24 \\
B & 06 08 47.0 &  20 38 36.4 &  7.51 (0.01) & 0.55 (0.01) & 3.89 (0.17) & 14.04 (0.23) & 1.29 (0.17) & 1.85 (0.09) & 0.71 (0.17) &  0.54 (0.09) & \nodata & \nodata \\
22........ & 06 08 45.9 &  21 31 49.5 &  1.89 (0.01) & 0.58 (0.01) & 3.35 (0.17) & 12.13 (0.23) & 1.56 (0.16) & 2.21 (0.09) & 0.64 (0.17) &  0.71 (0.09) & \nodata & \nodata \\
24........ & 06 08 49.5 &  21 33 41.4 &  2.66 (0.01) & 0.38 (0.01) & 1.71 (0.08) &  4.25 (0.11) & 0.51 (0.08) & 0.46 (0.04) & $<$ 0.87 & $<$ 0.48&0.32 (0.08) & 0.24 (0.05) \\
26........ & 06 08 53.6 &  21 38 33.8 &  3.03 (0.01) & 0.93 (0.01) & 2.82 (0.09) & 14.09 (0.12) & 1.63 (0.08) & 3.65 (0.05) & 1.30 (0.17) & 2.37 (0.10) & 0.36 (0.09) & 0.45 (0.05) \\
B & 06 08 52.4 &  21 37 57.2 &  3.51 (0.02) & 1.30 (0.02) & 1.75 (0.17) & 11.36 (0.23) & 1.13 (0.16) & 2.84 (0.09) & 1.02 (0.17) &  2.04 (0.10) & \nodata & \nodata \\
27........ & 06 08 53.6 &  20 29 25.5 &  8.35 (0.01) & 0.58 (0.01) & 3.21 (0.16) & 12.88 (0.22) & 1.04 (0.15) & 1.56 (0.08) & $<$ 0.79 & $<$ 0.44& \nodata & \nodata \\
28........ & 06 08 55.4 &  20 40 50.4 &  7.82 (0.03) & 0.55 (0.03) & 1.11 (0.16) &  2.82 (0.22) & $<$ 0.76 & $<$ 0.42 & $<$ 0.79 & $<$ 0.43 & \nodata & \nodata \\ 
31........ & 06 09 03.9 &  20 42 14.7 &  9.39 (0.03) & 0.39 (0.03) & 0.33 (0.06) &  0.88 (0.09) & $<$ 0.32 & $<$ 0.18 & $<$ 0.49 & $<$ 0.27 & $<$ 0.43 & $<$ 0.24 \\ 
32........ & 06 09 06.4 &  21 50 41.0 & -0.53 (0.01) & 1.27 (0.02) & 1.40 (0.08) &  8.47 (0.12) & 0.93 (0.08) & 2.24 (0.05) & 0.82 (0.16) & 1.54 (0.09) & $<$ 0.43 & $<$ 0.24 \\
34........ & 06 09 20.8 &  20 39 00.6 &  9.34 (0.04) & 0.64 (0.05) & 0.81 (0.14) &  2.08 (0.20) & 0.64 (0.15) & 0.62 (0.08) & $<$ 0.77 & $<$ 0.43& \nodata & \nodata \\
35........ & 06 09 30.7 &  21 23 35.3 & -1.00 (0.05) & 0.44 (0.06) & 0.34 (0.08) &  0.59 (0.12) & $<$ 0.44 & $<$ 0.25 & $<$ 0.47 & $<$ 0.26 & \nodata & \nodata \\ 
40........ & 06 10 51.9 &  20 29 33.7 &  7.83 (0.04) & 0.81 (0.04) & 0.94 (0.16) &  3.61 (0.22) & 0.61 (0.16) & 0.46 (0.09) & $<$ 0.85 & $<$ 0.47& \nodata & \nodata \\
B & 06 10 51.8 &  20 30 34.5 &  8.79 (0.02) & 0.52 (0.03) & 1.29 (0.16) &  2.87 (0.22) & $<$ 0.79 & $<$ 0.44 & $<$ 0.82 & $<$ 0.46 & \nodata & \nodata \\ 
42........ & 06 12 34.1 &  17 54 43.3 &  7.71 (0.02) & 0.50 (0.03) & 1.19 (0.15) &  3.04 (0.21) & 0.70 (0.15) & 0.63 (0.08) & $<$ 0.79 & $<$ 0.44& \nodata & \nodata \\
45........ & 06 12 53.3 &  18 00 14.3 &  8.66 (0.02) & 1.31 (0.02) & 2.28 (0.16) & 13.90 (0.22) & 1.46 (0.15) & 3.94 (0.09) & 1.23 (0.16) &  2.62 (0.09) & \nodata & \nodata \\
47........ & 06 12 54.1 &  17 59 31.1 &  7.20 (0.02) & 1.07 (0.02) & 1.82 (0.16) &  8.87 (0.22) & 1.34 (0.16) & 2.83 (0.09) & 1.04 (0.16) &  2.29 (0.09) & \nodata & \nodata \\
B & 06 12 55.9 &  17 57 42.4 &  6.25 (0.01) & 0.45 (0.01) & 2.51 (0.16) &  6.83 (0.22) & 0.94 (0.16) & 0.97 (0.09) & $<$ 0.83 & $<$ 0.47 & \nodata & \nodata \\
48........ & 06 13 11.0 &  17 58 46.6 &  5.86 (0.07) & 1.02 (0.07) & 0.98 (0.15) &  2.26 (0.21) & $<$ 0.76 & $<$ 0.42 & $<$ 0.81 & $<$ 0.45 & \nodata & \nodata \\ 
B & 06 13 08.8 &  17 58 21.6 &  6.62 (0.03) & 0.42 (0.03) & 0.95 (0.16) &  2.34 (0.22) & $<$ 0.78 & $<$ 0.43 & $<$ 0.82 & $<$ 0.45 & \nodata & \nodata \\ 
49........ & 06 13 29.3 &  17 55 33.2 &  7.85 (0.03) & 0.54 (0.03) & 0.94 (0.15) &  2.56 (0.21) & 0.70 (0.15) & 0.50 (0.08) & $<$ 0.80 & $<$ 0.44& \nodata & \nodata \\
B & 06 13 26.0 &  17 56 08.5 &  8.03 (0.01) & 0.29 (0.02) & 1.50 (0.15) &  2.41 (0.21) & $<$ 0.74 & $<$ 0.41 & $<$ 0.77 & $<$ 0.43 & \nodata & \nodata \\ 
C & 06 13 32.2 &  17 56 11.7 &  8.34 (0.03) & 0.37 (0.03) & 0.74 (0.15) &  1.63 (0.20) & $<$ 0.76 & $<$ 0.42 & $<$ 0.75 & $<$ 0.42 & \nodata & \nodata \\ 
50........ & 06 13 47.1 &  17 54 55.3 &  9.18 (0.01) & 0.44 (0.01) & 3.43 (0.16) & 11.20 (0.22) & 1.02 (0.16) & 1.08 (0.09) & $<$ 0.80 & $<$ 0.44& \nodata & \nodata \\
B & 06 13 45.4 &  17 55 10.9 &  9.22 (0.01) & 0.35 (0.01) & 2.73 (0.16) &  7.20 (0.22) & 0.85 (0.15) & 0.83 (0.08) & $<$ 0.76 & $<$ 0.42 & \nodata & \nodata \\
C & 06 13 49.6 &  17 54 29.4 &  9.63 (0.01) & 0.36 (0.01) & 2.63 (0.15) &  7.28 (0.20) & 0.75 (0.14) & 0.50 (0.08) & $<$ 0.76 & $<$ 0.42 & \nodata & \nodata \\
51........ & 06 13 59.3 &  17 52 52.8 &  8.94 (0.01) & 0.36 (0.01) & 2.01 (0.15) &  5.33 (0.20) & $<$ 0.73 & $<$ 0.41 & $<$ 0.73 & $<$ 0.41 & \nodata & \nodata \\ 
52........ & 06 14 09.7 &  17 43 50.3 &  6.65 (0.01) & 0.35 (0.02) & 1.52 (0.15) &  3.81 (0.21) & $<$ 0.76 & $<$ 0.42 & $<$ 0.79 & $<$ 0.44 & \nodata & \nodata \\ 
54........ & 06 14 24.9 &  17 45 03.9 &  7.90 (0.02) & 0.53 (0.02) & 1.69 (0.15) &  5.80 (0.20) & 0.88 (0.15) & 0.83 (0.08) & $<$ 0.75 & $<$ 0.41& \nodata & \nodata \\
B & 06 14 24.1 &  17 44 32.9 &  7.67 (0.01) & 0.61 (0.01) & 1.88 (0.15) &  6.85 (0.21) & 1.06 (0.14) & 1.46 (0.08) & 0.65 (0.15) &  0.48 (0.08) & \nodata & \nodata \\
C & 06 14 23.7 &  17 43 57.3 &  7.67 (0.04) & 0.47 (0.04) & 0.70 (0.15) &  1.10 (0.21) & $<$ 0.75 & $<$ 0.41 & $<$ 0.78 & $<$ 0.43 & \nodata & \nodata 
\enddata 
\tablecomments{Errors are given in parentheses.}
\tablenotetext{a}{B and C denote multiple ammonia pointings that fall within a single 1.1 mm source.}
\tablenotetext{b}{Upper-limits are $T_{mb} < 5\mbox{RMS}$ and $W < 5\mbox{RMS} \Delta v / \sqrt{N}$, where $\Delta v$ is the width of a single channel in velocity, and $N$ is the number of pixels over which the average RMS was calculated.}
\tablenotetext{c}{\nodata denotes sources that were not observed in the \ammonia(4,4) transition.}
\end{deluxetable}
\clearpage
\end{landscape}

\clearpage

\begin{deluxetable}{rccccccc}
\tabletypesize{\scriptsize}
\tablewidth{0pt}
\tablecaption{\label{gas_props}Derived Gas Properties}
\tablehead{
\colhead{ID} & \colhead{} & \colhead{T$_{\mbox{kin}}$\tablenotemark{b}} & \colhead{T$_{\mbox{ex}}$} & \colhead{$a$} & \colhead{$\sigma_{NT}$} & \colhead{H$_2$O} & \colhead{CCS}   \\                    
\colhead{Number\tablenotemark{a}} & \colhead{$\tau$(1,1)} & \colhead{(K)} & \colhead{(K)} & \colhead{(\kms)} & \colhead{(\kms)} & \colhead{maser?\tablenotemark{c}} & \colhead{emission?\tablenotemark{c}}    
}
\startdata
01........ & 1.56 (0.83) & $<$ 16.76 (1.42) &  3.50 (0.34) &  0.24(0.005) &  0.71 (0.11) & N & \nodata \\
03........ & 0.73 (0.35) & 17.15 (0.65) &  6.25 (1.40) &  0.24(0.002) &  0.58 (0.04) & N & \nodata \\
04........ & 0.96 (0.54) & $<$ 13.93 (1.00) &  4.87 (0.96) &  0.22(0.003) &  0.49 (0.06) & N & \nodata \\
05........ & 1.40 (0.38) & $<$ 12.49 (0.72) &  5.01 (0.50) &  0.21(0.003) &  0.48 (0.04) & N & \nodata \\
06........ & 1.26 (0.41) & 17.45 (0.72) &  3.88 (0.28) &  0.25(0.002) &  0.50 (0.04) & N & N \\
07........ & 1.43 (0.19) & 14.47 (0.32) &  5.17 (0.24) &  0.22(0.001) &  0.47 (0.02) & N & N \\
11........ & 0.40 (0.48) & 20.45 (0.93) &  5.81 (3.28) &  0.27(0.003) &  1.12 (0.11) & N & \nodata \\
12........ & 3.08 (0.26) & 13.39 (0.35) &  5.05 (0.16) &  0.22(0.001) &  0.41 (0.02) & N & \nodata \\
13........ & 1.68 (0.14) & 25.55 (0.22) &  6.53 (0.23) &  0.30(0.001) &  0.86 (0.02) & Y & \nodata \\
B & 0.85 (0.50) & 22.34 (0.86) &  5.02 (1.08) &  0.28(0.003) &  0.53 (0.05) & N & \nodata \\
16........ & 1.19 (0.12) & 18.93 (0.21) &  8.85 (0.47) &  0.26(0.001) &  0.77 (0.02) & Y & \nodata \\
17........ & 0.47 (0.19) & 25.49 (0.33) & 10.48 (2.73) &  0.30(0.001) &  0.91 (0.03) & N & \nodata \\
18........ & 1.20 (0.18) & 17.99 (0.32) &  6.83 (0.47) &  0.25(0.001) &  0.70 (0.03) & N & \nodata \\
19........ & 0.89 (0.30) & 19.48 (0.60) &  4.17 (0.38) &  0.26(0.002) &  0.90 (0.05) & N & N \\
20........ & 0.53 (0.27) & 19.26 (0.55) &  5.58 (1.24) &  0.26(0.002) &  0.75 (0.04) & N & N \\
21........ & 1.16 (0.11) & 17.54 (0.19) &  6.14 (0.24) &  0.25(0.001) &  1.00 (0.02) & N & N \\
B & 1.51 (0.16) & 17.04 (0.26) &  7.85 (0.38) &  0.24(0.001) &  0.54 (0.02) & N & \nodata \\
22........ & 1.29 (0.18) & 20.83 (0.30) &  7.58 (0.50) &  0.27(0.001) &  0.57 (0.02) & N & \nodata \\
24........ & 0.54 (0.21) & 16.44 (0.37) &  8.41 (1.89) &  0.24(0.001) &  0.37 (0.02) & Y & N \\
26........ & 0.50 (0.09) & 27.47 (0.19) & 11.61 (1.38) &  0.31(0.001) &  0.92 (0.02) & Y & N \\
B & 0.65 (0.22) & 28.57 (0.37) &  7.00 (1.25) &  0.32(0.001) &  1.30 (0.05) & Y & \nodata \\
27........ & 1.95 (0.17) & 16.04 (0.25) &  6.54 (0.22) &  0.24(0.001) &  0.57 (0.02) & N & \nodata \\
28........ & 0.76 (0.60) & $<$ 15.66 (1.09) &  4.86 (1.36) &  0.23(0.004) &  0.54 (0.07) & N & \nodata \\
31........ & 1.64 (0.85) & $<$ 19.15 (1.43) &  3.16 (0.23) &  0.26(0.005) &  0.38 (0.07) & N & N \\
32........ & 0.19 (0.14) & 29.26 (0.32) & 13.17 (7.34) &  0.32(0.001) &  1.26 (0.03) & Y & N \\
34........ & 0.71 (0.76) & 23.78 (1.36) &  4.22 (1.30) &  0.29(0.005) &  0.63 (0.09) & N & \nodata \\
35........ & 1.81 (1.33) & $<$ 18.19 (2.19) &  3.04 (0.34) &  0.25(0.008) &  0.43 (0.11) & N & \nodata \\
40........ & 0.64 (0.53) & 17.65 (1.02) &  4.96 (1.52) &  0.25(0.004) &  0.80 (0.09) & N & \nodata \\
B & 0.56 (0.49) & $<$ 15.40 (0.92) &  6.35 (2.73) &  0.23(0.003) &  0.51 (0.06) & N & \nodata \\
42........ & 0.47 (0.49) & 19.80 (0.90) &  6.75 (3.65) &  0.26(0.003) &  0.49 (0.05) & N & \nodata \\
45........ & 0.70 (0.17) & 29.55 (0.27) &  7.72 (1.00) &  0.32(0.001) &  1.30 (0.04) & Y & \nodata \\
47........ & 0.08 (0.00) & 32.27 (0.34) & 32.27 (0.34) &  0.34(0.001) &  1.06 (0.04) & Y & \nodata \\
B & 1.02 (0.26) & 19.02 (0.46) &  7.10 (0.89) &  0.26(0.002) &  0.44 (0.02) & N & \nodata \\
48........ & 0.05 (0.01) & $<$ 18.69 (1.43) & 18.69 (1.43) &  0.26(0.005) &  1.02 (0.14) & N & \nodata \\
B & 0.98 (0.80) & $<$ 18.44 (1.37) &  4.30 (0.99) &  0.25(0.005) &  0.41 (0.07) & N & \nodata \\
49........ & 1.22 (0.66) & 19.47 (1.15) &  4.03 (0.52) &  0.26(0.004) &  0.53 (0.07) & N & \nodata \\
B & 0.22 (0.54) & $<$ 15.72 (0.92) & 14.18 (9.50) &  0.23(0.003) &  0.28 (0.04) & N & \nodata \\
C & 0.49 (0.88) & $<$ 23.30 (1.48) &  5.35 (4.20) &  0.29(0.005) &  0.36 (0.07) & N & \nodata \\
50........ & 2.43 (0.18) & 14.21 (0.26) &  6.45 (0.19) &  0.22(0.001) &  0.43 (0.01) & N & \nodata \\
B & 2.06 (0.25) & 15.49 (0.36) &  6.06 (0.28) &  0.23(0.001) &  0.34 (0.02) & N & \nodata \\
C & 2.65 (0.25) & 13.11 (0.34) &  5.46 (0.19) &  0.21(0.001) &  0.35 (0.02) & N & \nodata \\
51........ & 2.60 (0.35) & $<$ 14.84 (0.48) &  4.69 (0.20) &  0.23(0.002) &  0.34 (0.02) & N & \nodata \\
52........ & 1.82 (0.46) & $<$ 15.54 (0.70) &  4.62 (0.34) &  0.23(0.002) &  0.34 (0.03) & N & \nodata \\
54........ & 0.80 (0.34) & 20.12 (0.60) &  6.16 (1.18) &  0.26(0.002) &  0.52 (0.04) & N & \nodata \\
B & 1.25 (0.27) & 23.20 (0.44) &  5.60 (0.46) &  0.28(0.002) &  0.60 (0.03) & N & \nodata \\
C & 1.77 (0.94) & $<$ 22.11 (1.48) &  3.47 (0.30) &  0.28(0.005) &  0.46 (0.08) & N & \nodata 
\enddata 
\tablecomments{Errors are given in parentheses.}
\tablenotetext{a}{B and C denote multiple ammonia pointings that fall within a single 1.1 mm source.}
\tablenotetext{b}{Upper-limits to T$_{\mbox{kin}}$ are due to a $< 5\sigma$ detection in the \ammonia(2,2) transition.}
\tablenotetext{c}{Either CCS or H$_2$O was observed, but not both.  \nodata marks sources not observed in the given transition.  Y denotes detected, N denotes not detected.}
\end{deluxetable}

\clearpage

\clearpage
\begin{deluxetable}{lccccccccc}
\tabletypesize{\scriptsize}
\tablewidth{0pt}
\tablecaption{\label{derived_masses_densities}Derived Masses and Densities}
\tablehead{
\colhead{ID} & \colhead{M$_{\mbox{iso}}$(120$^{\prime\prime}$)} & \colhead{M$_{\mbox{iso}}$(int)} & \colhead{M$_{\mbox{vir}}$(int)} & \colhead{n$_{p}$(int)} & \colhead{n$_{ex}$} & \colhead{$\Sigma$(int)} & \colhead{$N_{NH_3}$} & \colhead{$N_{H_2}$(int)} & \colhead{$X_{\ammonia}$} \\
\colhead{Number} & \colhead{(M$_{\odot}$)} & \colhead{(M$_{\odot}$)} & \colhead{(M$_{\odot}$)} & \colhead{($10^{3}$cm$^{-3}$)} & \colhead{($10^{3}$cm$^{-3}$)} & \colhead{($10^{-2}$ g cm$^{-2}$)} & \colhead{($10^{14}$ cm$^{-2}$)} & \colhead{($10^{21}$ cm$^{-2}$)} & \colhead{($10^{-8}$)} 
}
\startdata
01........ &  69(7) &  73(6) &      150(28) &  2.39(0.33) &  1.32(0.25) &  1.95(0.08) &  0.53(0.14) &  4.17(0.48) &  1.26(0.37) \\
03........ & 108(14) & 121(13) &       96(10) &  4.31(0.64) &  8.29(0.99) &  3.42(0.18) &  2.31(0.31) &  7.32(0.94) &  3.16(0.59) \\
04........ & 134(12) & 115(9) &       57(8) &  7.18(0.96) &  4.82(0.79) &  4.73(0.15) &  0.78(0.09) & 10.11(1.10) &  0.77(0.12) \\
05........ & 137(13) & 126(19) &       67(7) &  4.52(0.63) &  5.35(0.50) &  3.58(0.21) &  4.01(0.63) &  7.65(0.89) &  5.24(0.35) \\
06........ & 113(15) &  71(11) &       49(5) &  8.54(1.54) &  2.07(0.22) &  4.52(0.46) &  2.28(0.41) &  9.66(1.58) &  2.36(0.57) \\
07........ & 124(21) & 106(18) &       60(4) &  4.86(0.97) &  5.29(0.24) &  3.55(0.42) &  3.55(0.32) &  7.59(1.40) &  4.68(0.96) \\
11........ & 139(13) &  99(12) &      436(56) &  1.98(0.32) &  6.65(1.28) &  1.91(0.13) &  0.95(0.03) &  4.08(0.57) &  2.34(0.34) \\
12........ & 172(23) & 197(24) &       59(4) &  4.23(0.68) &  4.35(0.14) &  3.97(0.26) &  7.04(0.38) &  8.49(1.20) &  8.29(1.25) \\
13........ & 373(40) & 465(39) &      304(18) &  5.58(0.74) &  6.80(0.21) &  6.36(0.12) &  7.58(0.33) & 13.59(1.47) &  5.57(0.65) \\
16........ & 502(53) & 292(28) &      153(8) & 14.29(2.02) & 16.32(0.53) & 10.19(0.35) &  6.65(0.37) & 21.79(2.58) &  3.05(0.40) \\
17........ & 463(30) & 468(39) &      244(17) & 15.69(2.46) & 20.09(1.46) & 12.69(0.24) &  3.53(1.08) & 27.13(2.01) &  1.30(0.27) \\
18........ & 386(38) & 429(37) &      208(15) &  4.94(0.66) &  9.38(0.46) &  5.70(0.12) &  4.93(0.45) & 12.19(1.33) &  4.04(0.58) \\
19........ & 143(15) & 143(15) &      291(27) &  2.64(0.39) &  2.68(0.25) &  2.61(0.12) &  2.92(0.46) &  5.58(0.70) &  5.24(1.05) \\
20........ & 149(16) & 126(14) &      187(17) &  2.95(0.45) &  6.10(0.62) &  2.70(0.14) &  0.78(0.44) &  5.76(0.77) &  1.35(0.78) \\
21........ & 560(57) & 779(66) &      560(31) &  3.76(0.50) &  7.46(0.22) &  5.80(0.12) &  6.22(0.35) & 12.41(1.35) &  5.01(0.61) \\
22........ & 608(61) & 365(35) &      109(7) &  8.56(1.21) & 10.67(0.49) &  7.80(0.26) &  4.54(0.40) & 16.68(1.97) &  2.72(0.40) \\
24........ & 204(19) & 125(18) &       32(2) &  8.48(1.13) & 17.81(1.38) &  5.43(0.11) &  1.51(0.37) & 11.62(1.13) &  1.30(0.13) \\
26........ & 492(31) & 556(20) &      277(8) & 13.69(0.72) & 23.37(0.79) & 12.26(0.19) &  4.21(0.52) & 26.22(1.17) &  1.60(0.21) \\
27........ & 471(47) & 853(65) &      194(13) &  3.58(0.46) &  8.44(0.26) &  5.79(0.06) &  6.72(0.31) & 12.38(1.27) &  5.43(0.61) \\
28........ & 117(10) & 150(12) &      116(19) &  2.17(0.28) &  4.63(0.92) &  2.32(0.05) &  0.65(0.06) &  4.97(0.51) &  1.30(0.18) \\
31........ &  56(7) &  58(7) &       42(8) &  2.29(0.37) &  0.69(0.13) &  1.75(0.13) &  0.16(0.06) &  3.75(0.54) &  0.41(0.18) \\
32........ & 149(15) & 164(16) &      459(28) &  5.72(0.99) & 29.86(1.92) &  4.56(0.31) &  1.82(0.03) &  9.76(0.75) &  1.86(0.15) \\
34........ & 176(9) & 319(10) &      231(42) &  1.37(0.15) &  2.69(0.70) &  2.20(0.07) &  0.42(0.02) &  4.70(0.36) &  0.89(0.08) \\
35........ &  30(7) &  28(4) &       22(6) & 14.41(2.66) &  0.49(0.13) &  4.69(0.66) &  0.14(0.02) & 10.04(1.68) &  0.14(0.03) \\
40........ & 105(10) & 143(9) &      186(26) &  4.98(0.61) &  4.67(0.88) &  3.98(0.06) &  0.80(0.05) &  8.50(0.82) &  0.94(0.11) \\
42........ &  46(11) &  32(8) &       44(6) &  4.81(1.32) &  9.35(1.77) &  2.37(0.52) &  0.65(0.03) &  5.08(1.34) &  1.28(0.34) \\
45........ & 509(53) & 397(32) &      456(28) & 16.96(2.21) & 10.00(0.57) & 12.65(0.22) &  5.74(0.81) & 27.05(2.86) &  2.12(0.38) \\
47........ & 481(49) & 540(43) &      552(37) &  3.90(0.51) &   \nodata   &  5.26(0.08) &  2.23(0.03) & 11.25(1.18) &  1.98(0.21) \\
48........ &  84(9) &  87(8) &      310(53) &  2.67(0.38) &   \nodata   &  2.23(0.09) &  0.68(0.05) &  4.76(0.57) &  1.43(0.20) \\
49........ & 138(11) & 214(13) &      126(21) &  2.04(0.24) &  2.30(0.41) &  2.51(0.00) &  0.51(0.14) &  5.36(0.50) &  0.95(0.28) \\
50........ & 191(28) & 233(32) &       64(4) &  4.74(0.80) &  8.54(0.24) &  4.53(0.35) &  6.78(0.28) &  9.69(1.47) &  7.00(1.10) \\
51........ &  54(20) &  44(13) &       29(2) &  2.92(0.90) &  3.52(0.18) &  1.89(0.03) &  4.29(0.28) &  4.04(1.20) & 10.60(3.22) \\
52........ &  56(17) &  51(13) &       23(3) &  6.47(1.81) &  3.56(0.33) &  3.36(0.75) &  2.85(0.34) &  7.19(1.93) &  3.96(1.16) \\
54........ & 245(23) & 371(27) &      128(13) &  3.25(0.41) &  7.29(0.81) &  4.11(0.05) &  2.17(0.65) &  8.78(0.88) &  2.47(0.78) 
\enddata 
\tablecomments{Errors are given in parentheses.}
\end{deluxetable}
\clearpage

\clearpage
\begin{deluxetable}{lcccccccccl}
\tabletypesize{\scriptsize}
\tablewidth{0pt}
\tablecaption{\label{mean_properties}Statistical Summary}
\tablehead{
\colhead{} & \colhead{} & \multicolumn{4}{c}{Full Sample} & \colhead{} & \multicolumn{4}{c}{Subset 1}  \\ 
\cline{3-6} \cline{8-11} 
\colhead{Property}& \colhead{units} & \colhead{minimum} & \colhead{mean\tablenotemark{a}} & \colhead{median} & \colhead{maximum}  & \colhead{} & \colhead{minimum} & \colhead{mean\tablenotemark{a}} & \colhead{median} & \colhead{maximum} 
}
\startdata
$R_{maj}$ & pc &0.18 & 0.35(0.10) & 0.34 & 0.55 & & 0.22 & 0.37(0.10) & 0.35 & 0.55 \\
$R_{min}$ & pc & 0.13 & 0.23(0.05) & 0.23 & 0.37 & & 0.16 & 0.24(0.05) & 0.23 & 0.37  \\
$R_{obj}$ & pc & 0.20 & 0.56(0.19) & 0.51 & 0.99 & & 0.30 & 0.60(0.19) & 0.56 & 0.99  \\
\Snu(120\as) & Jy & 0.46 & 4.30(4.34) & 2.39 & 15.38 & & 0.80 & 5.49(4.51) & 3.85 & 15.38  \\
\Snu(int) & Jy &0.43 & 4.56(4.62) &  2.09 & 17.26 & & 0.56 & 5.84(4.78) &  4.65 & 17.26  \\
\tmb(1,1)  & K &0.33 & 1.76(0.96) &  1.69 &  4.15 & & 0.80 & 2.00(0.97) &  1.75 &  4.15 \\
\tmb(2,2)  & K & 0.27 & 0.82(0.41) &  0.69 &  1.72 & & 0.29 & 0.92(0.44) &  0.83 &  1.72 \\
\tex(1,1)  & K &  3.04 &   7.27(5.42) &  6.14 & 32.27 & &   3.88 &   7.78(5.60) &  6.45 & 32.27 \\
\tk  & K &  12.49 &  19.39(4.89) & 18.69 & 32.27 & &  13.39 &  20.56(5.08) & 19.47 & 32.27 \\
(\tk$-$\tex)  & K &  0.00 &  12.13(4.61) & 13.05 & 21.83 & &   0.00 &  12.78(4.48) & 13.24 & 21.83 \\
v$_{LSR}$  & \kms\ & -1.00 & 6.83(2.96) &  7.85 & 10.17 & & -0.53 & 6.93(2.76) &  7.85 & 10.12 \\
$\sigma_{v}$  & \kms\ &0.35 & 0.69(0.27) &  0.58 &  1.31 & & 0.38 & 0.75(0.27) &  0.71 &  1.31 \\
$a$  & \kms\ &  0.21 &   0.26(0.03) &  0.26 &  0.34 & &   0.22 &   0.27(0.03) &  0.26 &  0.34 \\
$\sigma_{NT}$  & \kms\ &  0.34 &   0.68(0.27) &  0.58 &  1.30 & &   0.37 &   0.74(0.27) &  0.70 &  1.30 \\
$\tau$(1,1)  & \nodata\ &  0.05 &   1.12(0.72) &  1.16 &  3.08 & &   0.08 &   1.02(0.70) &  0.80 &  3.08 \\
\miso(120\as)  & \msun\ &    31 &    230(180) &    150 &    610 & &     47 &    280(180) &    190 &    610 \\
\miso(int)  & \msun\ &    28 &    250(210) &    150 &    850 & &     33 &    300(220) &    230 &    850 \\
\mvir(int)  & \msun\ &    22 &    190(160) &    150 &    560 & &     33 &    220(160) &    190 &    560 \\
\mvir/\miso (int)  & \nodata\ &  0.23 &   1.00(0.95) &  0.70 &  4.37 & &   0.23 &   0.96(0.93) &  0.65 &  4.37 \\
\np(int)  & 10$^3$ \cmv\ &  1.37 &   5.91(4.29) &  4.74 & 16.96 & &   1.37 &   6.23(4.42) &  4.81 & 16.96 \\
$\Sigma$(int)  & g \cmc\ &  0.018 &   0.048(0.030) &   0.041 &   0.127 & &   0.019 &   0.055(0.032) &   0.045 &   0.127 \\
$n_{ex}$  & $10^3$ \cmv\ &   0.49 &    7.96(6.79) &    6.65 &   29.90 & &    2.07 &    9.60(7.05) &    8.29 &   29.90 \\
\np (int)/$n_{ex}$  & \nodata\ &   0.03 &    1.50(1.00) &    1.25 &    5.22 & &    0.24 &    1.73(1.00) &    1.80 &    5.22 \\
$N_{NH_3}$  & $10^{13}$ \cmc\ &   1.41 &   29.68(23.61) &   23.13 &   75.76 & &    4.18 &   34.74(23.93) &   29.24 &   75.76 \\
$N_{H_2}$(int)  & $10^{21}$ \cmc\ &   3.75 &   10.27(6.52) &    8.78 &   27.13 & &    4.08 &   11.71(6.94) &    9.69 &   27.13 \\
$N_{NH_3}/N_{H_2}$(int)  & $10^{-8}$ &   0.14 &    3.00(2.41) &    2.34 &   10.60 & &    0.89 &    3.08(2.04) &    2.36 &    8.29 
\enddata 
\tablenotetext{a}{Standard deviation is given in parentheses.}
\end{deluxetable}
\clearpage

\end{document}